\documentclass[twocolumn]{aastex701}

\usepackage{CJKutf8}
\usepackage{amsmath}	
\usepackage{threeparttable}

\newcommand{\project}[1]{\textsl{#1}}

\newcommand{\gaia}{\project{Gaia}}

\newcommand{\T}{\ensuremath{\mathrm{T}}}

\newcommand{\appropto}{\mathrel{\vcenter{
  \offinterlineskip\halign{\hfil$##$\cr
    \propto\cr\noalign{\kern2pt}\sim\cr\noalign{\kern-2pt}}}}}

\graphicspath{{./}{figures/}}

\begin{document}

\title{Bernhard-1: An Eccentric Binary Periodically Obscured by its Misaligned Circumbinary Disk}

\correspondingauthor{Zhecheng Hu, Wei Zhu}
\email{hzc22@mails.tsinghua.edu.cn}
\correspauthortrue
\email{weizhu@tsinghua.edu.cn}

\author[orcid=0009-0000-6461-5256]{Zhecheng Hu (\begin{CJK*}{UTF8}{gbsn}胡哲程\end{CJK*})}
\affiliation{Department of Astronomy, Tsinghua University, Beijing 10084, China}
\email{hzc22@mails.tsinghua.edu.cn}

\author[orcid=0000-0003-4027-4711]{Wei Zhu (\begin{CJK*}{UTF8}{gbsn}祝伟\end{CJK*})}
\affiliation{Department of Astronomy, Tsinghua University, Beijing 10084, China}
\email{weizhu@tsinghua.edu.cn}

\author[orcid=0000-0003-0853-6427]{Ping Chen (\begin{CJK*}{UTF8}{gbsn}陈平\end{CJK*})}
\affiliation{Institute for Advanced Study in Physics, Zhejiang University, Hangzhou 310058, People's Republic of China}
\email{ping.chen@zju.edu.cn}

\author{Richard Post}
\affiliation{Post Observatory, Lexington, MA, USA}
\email{rspostrspost@gmail.com}

\author[orcid=0000-0001-6000-3463]{Weicheng Zang (\begin{CJK*}{UTF8}{gbsn}臧伟呈\end{CJK*})}
\affiliation{Department of Astronomy, Westlake University, Hangzhou 310030, Zhejiang Province, China}
\email{zangweicheng@westlake.edu.cn}

\begin{abstract}
Bernhard-1 is a proposed KH 15D-like circumbinary disk occultation (CBO) system, but its binary nature and disk geometry have not previously been confirmed. We present new optical and near-infrared spectroscopy together with multi-band photometric monitoring of the system. The radial velocities confirm that Bernhard-1 hosts a highly eccentric binary with $e = 0.80 \pm 0.09$, confirming that the periodic photometric variability arises from occultation by a misaligned circumbinary disk. Joint modeling of the spectra and phase-dependent spectral energy distributions yields pre-main-sequence components with masses of $\sim 1.1\,M_\sun$ and $\sim 0.8\,M_\sun$. Combining stellar isochrones with the measured lithium abundance yields a system age of $\sim$ 10 Myr. Together with the spatial, astrometric, and metallicity properties of Bernhard-1, this suggests that Bernhard-1 is probably a member of the open cluster Dolidze 42. By combining the RV orbit with a semi-transparent occultation-screen model, we infer a disk--binary mutual inclination of roughly $50^\circ$ or $130^\circ$, with the degeneracy arising from the unknown disk rotation direction. This geometric method can be applied to any CBO system once radial velocity monitoring yields an orbital solution. The new light curves deviate from earlier model predictions, consistent with ongoing disk precession, while the phase-dependent H$\alpha$ profiles indicate pulsed accretion near periastron. Bernhard-1 therefore joins KH 15D and Bernhard-2 as a rare spectroscopically confirmed CBO system.
\end{abstract}

\keywords{Circumstellar disks; Variable stars; Spectroscopy; Pre-main sequence stars}

\section{Introduction} \label{sec:intro}

Binary stars are common outcomes of star formation, with a multiplicity fraction of nearly 50\% among field solar-type stars in the solar neighborhood \citep{Raghavan10_survey_stellar, Duchene13_stellar_multiplicity}. Planets orbiting both stars in a binary, known as circumbinary planets (CBPs), therefore provide an important test of planet formation and orbital evolution in dynamically complex environments. To date, 17 CBPs have been confirmed, with 14 found by transit \citep[e.g., ][]{Doyle11_kepler-16_transiting, Kostov20_toi-1338_tess} and the rest by eclipse timing variations \citep[][]{Goldberg23_5m_sub} or by a dedicated radial velocity survey such as the Binaries Escorted By Orbiting Planets (BEBOP) survey \citep{Martin19_bebop_radial-velocity}.

Of particular interest is the alignment between the binary and planetary orbital planes. Transit detections, however, are biased toward coplanar systems, and all known transiting CBPs have mutual inclinations $<4.5^\circ$ \citep{Martin14_planets_transiting, Chen22_number_transits}. Yet highly misaligned and even polar CBPs are predicted to exist, since the polar configuration is another stable attractor, especially for eccentric binaries \citep[e.g., ][]{Martin17_polar_alignment, Zanazzi18_inclination_evolution, Childs21_formation_polar}. Demographic arguments based on the transiting sample suggest that, if giant planets occur with similar frequencies around single and binary stars, the observed yield is consistent with a coplanar CBP population \citep{Armstrong14_abundance_circumbinary,Martin14_planets_transiting}. However, this inference remains indirect because highly inclined systems are intrinsically difficult to detect. The true mutual-inclination distribution of CBPs therefore still lacks direct observational constraints.

Because circumbinary protoplanetary disks define the birth planes of circumbinary planets, their alignment distribution offers a useful proxy for the primordial inclination distribution of CBPs. Many such disks are observed to be misaligned with eccentric central binaries \citep[e.g., ][]{Kearns98_additional_periodic, Kohler11_orbit_gg, Kennedy12_99, Kennedy19_circumbinary_protoplanetary, Brinch16_misaligned_disks, Lacour16_m-dwarf_star, FernandezLopez17_strongly_misaligned, Kenworthy22_eclipse_v773,Hu24_eccentric_binary}. This misalignment appears especially common around binaries with intermediate orbital periods \citep[30--$10^5$ days;][]{Czekala19_degree_alignment}, which are longer than those of the currently known transiting CBP hosts. Misaligned circumbinary disks therefore constrain both the initial conditions of CBP evolution and a binary parameter space largely inaccessible to the present CBP sample.

Among misaligned circumbinary disks, the circumbinary disk occultation (CBO) systems, of which KH 15D is the prototype, are uniquely valuable \citep{Kearns98_additional_periodic}. In this configuration, a misaligned circumbinary disk periodically occults the central binary, producing characteristic flux variations \citep{Chiang04_circumbinary_ring, Winn04_kh_15d,Winn06_orbit_occultations}. Decades of photometric follow-up of KH 15D have revealed long-term changes in its light curve driven by disk precession, and detailed modeling has constrained the disk--binary mutual inclination to $\sim 15^\circ$ and the disk warp to $\sim 10^\circ$ \citep{Winn06_orbit_occultations,Poon21_constraining_circumbinary}. Such systems are therefore rare cases in which time-domain photometry, when combined with orbital information, can be converted into geometric constraints on a young circumbinary disk.

Large-scale photometric surveys are now rapidly expanding the CBO sample, but spectroscopic confirmation lags far behind. More than 30 CBO candidates have been proposed from the Zwicky Transient Facility \citep[ZTF, ][]{Bellm19_zwicky_transient,Masci19_zwicky_transient,Zhu2022_Two,Hu26_six}, the Optical Gravitational Lensing Experiment \citep[OGLE, ][]{Udalski97_optical_gravitational, Udalski03_optical_gravitational, Udalski15_ogle-iv_fourth,Urbanowicz26_thirty_circumbinary}, All-Sky Automated Survey for Supernovae \citep[ASAS-SN, ][]{Shappee2014_MAN, Kochanek2017_All-Sky, Fores-Toribio2025_ASASSN-24fw, Zakamska2025_ASASSN-24fw} and the VISTA Variables in the Via Lactea survey \citep[VVV, ][]{Minniti10_vista_variables, Lucas24_most_variable}. Yet only one, Bernhard-2, has had its central binary confirmed by the radial velocity (RV) method \citep{Hu24_eccentric_binary}, with $e \simeq 0.7$ and a potentially highly misaligned disk.

In this work, we confirm Bernhard-1, which is the other candidate CBO system identified in \citet{Zhu2022_Two}, based on dedicated spectroscopic and multi-band photometric observations. Located at $(20^{\mathrm h}20^{\mathrm m}55.22^{\mathrm s},+38^\circ13^\prime23.1^{\prime\prime})$, Bernhard-1 is a 192-day system with $r$-band magnitudes of 17.30 outside occultation and 19.42 during occultation, making it both optically fainter and longer-period than Bernhard-2 \citep{Zhu2022_Two}.  The radial velocities reveal a highly eccentric central binary, confirming that the periodic variability arises from occultation by circumbinary material. We jointly model the spectra and spectral energy distributions (SEDs) to infer the stellar properties, and we model the light curve with a semi-transparent occultation screen. Combining the screen geometry with the RV orbit yields a rough disk--binary mutual inclination of $50^\circ$ or $130^\circ$. The light curve also shows long-term changes attributable to disk precession, while the phase-dependent H$\alpha$ profiles indicate pulsed accretion near pericenter passage. All of these features closely resemble the observations of KH 15D and Bernhard-2.

The paper is organized as follows. Section~\ref{sec:obs} describes our spectroscopic and photometric observations. Section~\ref{sec:st-mod} infers the stellar parameters from the spectra and SEDs. Section~\ref{sec:model-rv} models the RV orbit, while Section~\ref{sec:model-lc} presents our semi-transparent occultation-screen model. Section~\ref{sec:discussion} discusses the cluster association, lithium age, disk geometry, precession, and pulsed accretion. We summarize our conclusions in Section~\ref{sec:conclusion}.

\section{Observations and Data Reduction} \label{sec:obs}

\begin{figure*}[ht!]
\centering
\includegraphics[width=0.8\textwidth]{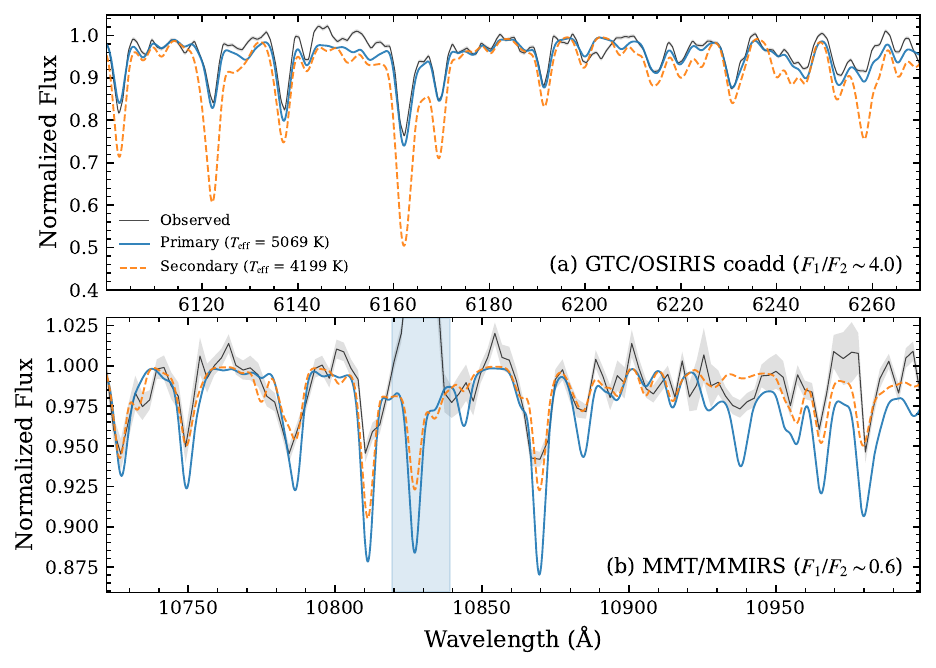}
\caption{Comparison between the combined GTC/OSIRIS spectra, obtained out of occultation, and the MMT/MMIRS spectrum, obtained at the end of ingress. The upper panel shows the combined GTC spectrum, in which the flux ratio between the primary and the secondary $F_1 / F_2$ in the corresponding wavelength range is about 4.0. The lower panel shows the MMT/MMIRS observation, which was obtained when $F_1 /F_2 \sim 0.6$ in the corresponding wavelength. The black curves and shaded bands indicate the observed spectra and the corresponding uncertainties. The blue solid line and the orange dashed line show the synthesized stellar spectra using the best-fit stellar parameters in Table \ref{tab:star-prop} for the primary and secondary star, respectively. The blue shaded rectangle in the lower panel indicates the He I 10830 line, which shows emission and does not align well with either templates.}
\label{fig:spec-comp}
\end{figure*}

\subsection{GTC/OSIRIS} \label{subsec:obs-gtc}

We obtained seven spectra of Bernhard-1 between 2025 June 18 and August 21 with the OSIRIS instrument \citep{Cepa2010_GTC_OSIRIS} installed on the 10.4\,m Gran Telescopio Canarias (GTC). Observations were taken at random orbital phases during the out-of-occultation windows.
We used the long-slit mode of OSIRIS, with the R2500R grism and a slit width of 0.6$\arcsec$ for Bernhard-1. This yields a wavelength coverage from 5575 ${\rm \AA}$ to 7685 ${\rm \AA}$ and a spectral resolution of $\sim 2500$. The exposure time is 900\,s for Bernhard-1.

The GTC data were reduced with the \texttt{PypeIt}\footnote{\url{https://pypeit.readthedocs.io/en/latest/}} package \citep{pypeit:joss_pub,pypeit:zenodo} following the standard process. We refined the wavelength solution after the lamp calibration by calibrating the telluric emission lines to the sky spectrum model, following \citet{Hu24_eccentric_binary}. Specifically, we extracted the observed sky emission lines with \texttt{PypeIt} and compared them to the wavelengths of the sky spectrum model, available on the GTC website \footnote{\url{https://www.gtc.iac.es/instruments/osiris+/media/sky/sky_res2500.txt}}. Applying the same procedure of \citet{Hu24_eccentric_binary}, we obtained a per-resolution wavelength uncertainty of $\sim 0.2 \, \rm \AA$ based on the internal scatter. At $\sim6500,\rm \AA$, with more than 30 spectral lines contributing to the RV measurement, this wavelength uncertainty corresponds to a velocity uncertainty of $\sim 1$ to $2\,\rm km,s^{-1}$ \citep[e.g.,][]{Bouchy2001}. The sixth GTC observation has a relatively larger wavelength uncertainty of $\sim 0.36 \, \rm \AA$, and thus is assigned a larger RV noise during modeling.

\subsection{MMT/MMIRS} \label{subsec:obs-mmt}

We obtained two spectra of Bernhard-1 with the MMIRS instrument \citep{Mcleod2012_MMT_MMIRS} on the 6.5\,m MMT Telescope on 2025 September 10 and November 3. The two observations were conducted during the ingress of the occultation as well as the occultation itself. We used the zJ filter with the J grism, with a 0.6~$\arcsec$ slit width in the long-slit mode. This yields a wavelength coverage from 0.9 $\rm \mu m$ to 1.5 $\rm \mu m$ and a spectral resolution of $\sim 2000$. The exposure times are 7200\,s for each observation. Note that the ingress observation has the low noise gain configuration (gain=0.95), while the occultation observation has the high dynamical range gain configuration (gain=2.68)\footnote{See the MMIRS Observer's Manual:
\url{https://lweb.cfa.harvard.edu/mmti/mmirs/MMIRSObsManual.pdf}.}.

The MMIRS data were reduced using two pipelines because the official data product for the second observation exhibited strong systematics. In addition to the standard product from the SAO Telescope Data Center using the IDL pipeline \citep{Chilingarian15_data_reduction}, we developed a customized pipeline \footnote{\url{https://github.com/zhechenghu/mmt-mmirs-up-the-ramp-pypeit}}. The two pipelines yield generally consistent results, except that our custom pipeline yields fewer outliers due to a different up-the-ramp fitting algorithm. We therefore use the customized pipeline for further reductions. In this pipeline, we combined the up-the-ramp fitting code \texttt{fitramp}\footnote{\url{https://github.com/t-brandt/fitramp}} \citep{Brandt24_optimal_fitting,Brandt24_likelihood-based_jump} with the \texttt{PypeIt} package. While our code, which wraps \texttt{fitramp}, yields ramp-fitted and overscan-corrected dark, flat, and science images, \texttt{PypeIt} handles with the standard calibrations for a typical long-slit spectrum.

While the reductions yield SNRs of about 150 and 70 for the ingress and occultation observations, we found that the observation during occultation with gain=2.68 suffers severely from systematics. The median absolute deviations (MADs) for the two observations are about two and seven times larger than those derived from the pipeline-produced noise, respectively, which means that the observations during occultation are severely affected by systematics. Therefore, we only use the MMIRS observation during ingress for further modeling.

\subsection{Photometric Observations} \label{subsec:obs-phot}

We have obtained photometric observations from the ZTF as well as the Post Observatory. For the ZTF data, we retrieved the ZTF $g$, $r$, and $i$ band light curve from ZTF DR24 after JD 2460800, within a range of 1.5 arcsec from Bernhard-1. We retained only observations with good quality flags, i.e., \texttt{catflags} $<$ 32768 \citep{Masci19_zwicky_transient}.

We performed follow-up photometric observations of Bernhard-1 using the telescopes operated by Post Observatory (hereafter PO) from 2024 July 14 to 2025 September 9. The images were taken with a Corrected Dall-Kirkham 17-inch telescope (hereafter CDK17), a Corrected Dall-Kirkham 24-inch telescope (hereafter CDK24), and a Ritchey-Chretien 32-inch telescope (hereafter RC32). The CDK17 telescope is located at Trenton, Maine, USA, and both the CDK24 and RC32 telescopes are located at Mayhill, New Mexico, USA. 

Basic image reductions, including bias subtraction, dark subtraction, and flat fielding, have been performed with the MaxIm DL Version 6.50 software. We performed point-spread function (PSF) photometry for Bernhard-1 and the reference objects in the field of each image using DoPHOT \citep{Schechter1993_DoPHOT, AlonsoGarcia2012_DoPHOT_C}. The photometric calibration is against the SDSS magnitudes transformed from the Pan-STARRS photometric catalog \citep{Flewelling2020_PS1}. All the photometry procedures follow the method outlined in \cite{Chen2022}.

\section{Stellar Parameter Modeling} \label{sec:st-mod}

\subsection{Spectral Fitting} \label{subsec:spec-fit}

We model each spectrum with a single-star template, because one star dominates the observed flux in each observation. Outside occultation, the secondary contributes only $\sim 0.16$ of the total flux in the wavelength range of the low-resolution GTC spectra, similar to the case of Bernhard-2 \citep{Hu24_eccentric_binary}, so we use the GTC/OSIRIS spectra to derive the stellar parameters and RVs of the primary star. The MMT/MMIRS spectrum, in contrast, was obtained during ingress, when the primary contributes about one-third of the total flux. We therefore use its single-star fit only as a consistency check rather than as an unbiased measurement, noting also that its RV may be distorted by a line-profile effect during partial occultation analogous to the Rossiter-McLaughlin effect \citep{Rossiter1924,McLaughlin1924}, which is believed to affect RV measurement of KH 15D \citep{Winn06_orbit_occultations}.

We fit the spectra with \texttt{starfish} \citep{Czekala2015_Constructing}, which interpolates synthetic spectral grids and models correlated residuals with a Gaussian Process (GP), reducing the biases in the inferred stellar parameters caused by imperfect templates and observational systematics. Separate optical and NIR emulators were trained for the GTC and MMT spectra, covering 588--684 nm and 996--1350 nm, respectively. The underlying grids were generated with \texttt{iSpec} \citep{Blanco-Cuaresma2019_iSpec2}, using the SPECTRUM radiative-transfer code \citep{Gray1994_Spectrum}, ATLAS9 model atmospheres \citep{Castelli2003_ATLAS9}, the \citet{Grevesse1998SS_solar_composition} solar abundance scale, and the VALD linelist \citep{Ryabchikova2015_VALD3}. Both grids span $T_{\rm eff}=$4000--6000 K, $\log g=$3.0--5.0, and ${\rm [Fe/H]}=-1.5$ to 0.5, with spacings of 200 K, 0.5 dex, and 0.3 dex, respectively. Microturbulent and macroturbulent velocities were estimated using the empirical prescriptions implemented in \texttt{iSpec}. Although the young stars like Bernhard-1 may rotate more rapidly, line broadening in our spectra is dominated by the low spectral resolution. As a result, we fix the projected rotational velocity to 1.60 km s$^{-1}$ and the limb-darkening coefficient to 0.6 \citep{Gray2005_OASP}. Because our grids are continuum-flattened, we removed the Planck scaling term from the GP covariance matrix, which is designed for flux-calibrated grids such as the PHOENIX library \citep[e.g.,][]{Husser2013_phoenix}.

The optical and NIR spectra were fit with slightly different sampling strategies. For the optical spectra, we adopted uniform priors within the grid for $T_{\rm eff}$ and ${\rm [Fe/H]}$, and a prior on $\log g$ from an initial SED fitting (Section \ref{subsec:sed-fitting}), because low-resolution spectra alone constrain surface gravity only weakly \citep{Xiang15_lamost_stellar}. We first optimized the third-order Chebyshev baseline, stellar parameters, and GP hyper-parameters with the Nelder-Mead method \citep{Nelder1965} implemented in \texttt{scipy} \citep{Virtanen2020_SciPy} for all the optical spectra. We then fixed the GP hyper-parameters and sampled the posterior around the best fit with \texttt{emcee} for the seven spectra \citep{Foreman-Mackey2013_emcee}. For the single usable NIR spectrum, we directly sampled the posterior with free GP hyper-parameters.

We adopt the GTC-derived stellar parameters and RVs with additional systematic noise floors to downstream modeling. As low-resolution spectra are known to suffer from systematic uncertainties, we added noise floors of 100 K in $T_{\rm eff}$, 0.3 dex in $\log g$, 0.1 dex in metallicity. An RV noise floor of 1~km/s will be added as a jitter term in RV modeling process. These noise values are comparable to the reported uncertainties of high-SNR LAMOST spectra \citep{Xiang15_lamost_stellar} and previous GTC/OSIRIS stellar spectra \citep[e.g., ][]{Aguado21_s2_stream}. The best-fit primary parameters, $T_{\rm eff,1} \approx 5100$ K with the corresponding $\log g_1$ and $\rm [Fe/H]$, are used as a prior in the SED fitting, and the RV measurements are listed in Table \ref{tab:rv-data}.

\begin{table}
\centering
\caption{GTC/OSIRIS RV data of Bernhard-1 used in the orbit modeling. Here BJD$^\prime=$BJD$-2460000$. The RV uncertainties include the 1 km s$^{-1}$ RV noise floor.}
\label{tab:rv-data}
{\footnotesize
\begin{tabular}{lccc}
\hline\hline
Date (UTC) & BJD$^\prime$ & RV (km s$^{-1}$) & $\sigma_{\mathrm{RV}}$ (km s$^{-1}$) \\
\hline
2025-06-18 & 844.613 & -6.03 & 1.74 \\
2025-06-25 & 851.557 & 4.73 & 1.68 \\
2025-07-04 & 860.610 & 25.28 & 1.82 \\
2025-07-13 & 870.464 & 9.13 & 2.23 \\
2025-07-28 & 885.497 & -10.59 & 1.69 \\
2025-08-11 $^{\rm a}$ & 899.446 & -17.46 & 5.10 \\
2025-08-21 & 908.516 & -5.87 & 1.60 \\
\hline
\hline
\end{tabular}
}
\begin{tablenotes}
\item[a] $^{\rm a}$ The observation with relatively worse wavelength uncertainty.
\end{tablenotes}
\end{table}

The phase dependence of the observed spectra provides evidence for the binary nature of Bernhard-1 independent of the RV measurements. While the GTC and MMIRS metallicities agree within 1$\sigma$, the MMIRS spectrum yields a substantially cooler effective temperature of $\sim 4400$ K than the GTC value of $\sim 5100$ K. Figure \ref{fig:spec-comp} compares the combined out-of-occultation GTC spectrum with the MMIRS spectrum obtained near the end of ingress. The warmer primary-star template matches the GTC spectrum better, whereas the cooler secondary-star template better matches the MMIRS spectrum, in which the secondary contributes the majority of the observed flux.

\begin{table}
\centering
\begin{threeparttable}
\caption{The adopted SED, spectral, RV, and static-screen best-fit parameters for Bernhard-1, as well as inferred properties marked with $^*$. The subscripts $1$ and $2$ indicate the primary and secondary star in the binary, respectively. Here BJD$^\prime=$BJD$-2460000$.}
\label{tab:star-prop}
{\footnotesize
\setlength{\tabcolsep}{3pt}
\begin{tabular*}{\columnwidth}{ccc}
\hline\hline
Parameter & Symbol & Value \\
\hline
\multicolumn{3}{l}{\bf{SED and spectral fitting}} \\
S1 EEP & EEP$_{1}$ & $175 \pm 4$ \\
S2 EEP & EEP$_{2}$ & $165 \pm 5$ \\
Log of age (yr) & $\log \mathrm{Age}$ & $6.99 \pm 0.13$ \\
Metallicity $^{\rm a}$ & $\mathrm{[Fe/H]}$ & $-0.22 \pm 0.10$ \\
Distance & $d$ (kpc) & $1.20 \pm 0.12$ \\
Extinction & $A_V$ (mag) & $3.17 \pm 0.14$ \\
S1 Effective Temp. $^{\rm a}$ & $T_{\mathrm{eff},1}$ (K) & $5050 \pm 120$ \\
S1 Surface Gravity $^{\rm a}$ & $\log g_{1}$ (cgs) & $4.27 \pm 0.06$ \\
S1 Mass & $M_{1}^{*}$ ($M_{\odot}$) & $1.11 \pm 0.09$\tnote{a} \\
S1 Radius & $R_{1}^{*}$ ($R_{\odot}$) & $1.28 \pm 0.14$ \\
S2 Effective Temp. & $T_{\mathrm{eff},2}^{*}$ (K) & $4200 \pm 90$ \\
S2 Surface Gravity & $\log g_{2}^{*}$ (cgs) & $4.29 \pm 0.06$ \\
S2 Mass & $M_{2}^{*}$ ($M_{\odot}$) & $0.82 \pm 0.06$ \\
S2 Radius & $R_{2}^{*}$ ($R_{\odot}$) & $1.07 \pm 0.10$ \\
\hline
\multicolumn{3}{l}{\bf{RV fitting}} \\
Orbital period (fixed) & $P$ (d) & $191.41$ (fixed) \\
Periastron time & $T_{P}$ (BJD$^\prime$) & $863.9^{+0.9}_{-3.3}$ \\
Eccentricity & $e$ & $0.80^{+0.09}_{-0.08}$ \\
Argument of periastron & $\omega_{1}$ (deg) & $4^{+4}_{-10}$ \\
RV semi-amplitude & $K_{1}$ (km s$^{-1}$) & $27^{+29}_{-7}$ \\
System velocity & $\gamma$ (km s$^{-1}$) & $-5.8^{+1.3}_{-1.5}$ \\
Binary Separation & $a_{\rm b}^{*}$ (AU) & $0.79^{+0.07}_{-0.03}$ \\
Inclination estimate & $i_{\rm b}^{*}$ (deg) & $54^{+8}_{-12}$ \\
\hline
\multicolumn{3}{l}{\bf{Light-curve / occultation fitting}} \\
Screen angle & $\theta_0$ (deg) & $128.94 \pm 0.08$ \\
Screen offset & $|d_0|$ (AU) & $0.4094 \pm 0.0004$ \\
Opacity scale & $s_0$ (AU) & $0.03336 \pm 0.00018$ \\
\hline
\hline
\end{tabular*}
}
\begin{tablenotes}
\item[a] These quantities are constrained by both the GTC spectral and SED fitting.
\end{tablenotes}
\end{threeparttable}
\end{table}

\subsection{SED Fitting} \label{subsec:sed-fitting}

Following \citet{Hu24_eccentric_binary} and \citet{Hu26_six}, we constrain the physical parameters of both stars by simultaneously fitting the out-of-occultation and in-occultation SEDs, which measure the total flux of the binary and the flux of the secondary, respectively. The SED model is built on the \texttt{isochrones} package \citep{Morton2015_isochrones}, which interpolates the MESA Isochrones and Stellar Tracks \citep[MIST,][]{Choi2016_MIST,Dotter2016_MIST} to predict broadband magnitudes. The magnitude of a single star in band $B$ is given by
\begin{equation}
m_{B} = f( {\rm EEP}, \log {\rm age}, [{\rm Fe/H}], d, A_V)
\end{equation}
where EEP is the equivalent evolutionary phase \citep{Dotter2016_MIST}, $d$ is the distance, and $A_V$ is the $V$-band extinction, for which we adopt the \citet{Cardelli1989_extinction} law with $R_V = 3.1$. Assuming the two stars are coeval, they share all parameters except their EEPs, giving six free parameters in total: $({\rm EEP}_1, {\rm EEP}_2, \log {\rm age}, [{\rm Fe/H}], d, A_V)$. The model out-of-occultation magnitude in each band is the sum of the two stellar fluxes, while the in-occultation magnitude is that of the secondary alone.

The SED fitting is performed on the photometry compiled in \citet{Zhu2022_Two}, together with the \gaia\ DR3 parallax as a Gaussian prior on the distance. A noise floor of 0.03~mag is added in quadrature to all photometric uncertainties to account for calibration systematics and imperfections in the stellar models. The WISE W3 and W4 measurements are excluded because they show clear infrared excess. We first locate the maximum-likelihood solution by optimizing the total log-likelihood, and then sample the posterior with \texttt{emcee} \citep{Foreman-Mackey2013_emcee}.

\begin{figure}[ht!]
\centering
\includegraphics[width=0.45\textwidth]{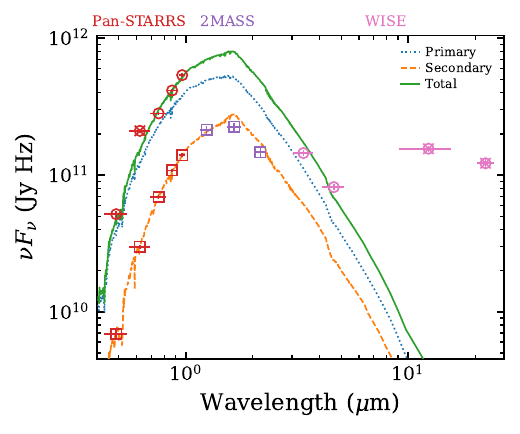}
\caption{Photometric measurements and the best-fit SED models of Bernhard-1. The red, purple, and pink markers represent data from Pan-STARRS, 2MASS, and WISE, respectively. Circles show the out-of-occultation photometry, fit by the total, i.e., primary plus secondary, model SED, and squares show the in-occultation photometry, fit by the secondary-only model SED. The WISE W3 and W4 measurements are excluded from the fit due to infrared excess and marked with crosses.}
\label{fig:sed}
\end{figure}

Because the SED and spectral analyses constrain complementary parameters, we iterated between them to get a converged solution. The $\log g_1$ posterior from an initial SED run was adopted as the prior for the GTC spectral fitting as described in Section~\ref{subsec:spec-fit}, because low-resolution spectra alone constrain it only weakly. The spectroscopic $T_{\rm eff,1}$, $\log g_1$, and $[{\rm Fe/H}]$ of the primary, with the systematic noise floors described in Section~\ref{subsec:spec-fit}, were then applied as priors on the second, final SED run. The best-fit parameters and uncertainties are listed in Table~\ref{tab:star-prop}, and the corresponding model SEDs are compared with the photometry in Figure~\ref{fig:sed}.

As an independent consistency check, we compare the derived secondary temperature with the MMIRS spectrum fitting result. Although the MMIRS spectrum was taken during ingress and is thus contaminated by the primary, which still contributes $\sim 1/3$ of the flux, its fitted temperature of $\sim$4400~K is substantially lower than that of the primary and reasonably close to the derived $T_{\rm eff,2} = 4200 \pm 90$~K, supporting the SED decomposition.

\begin{figure*}[ht!]
\centering
\includegraphics[width=1.0\textwidth]{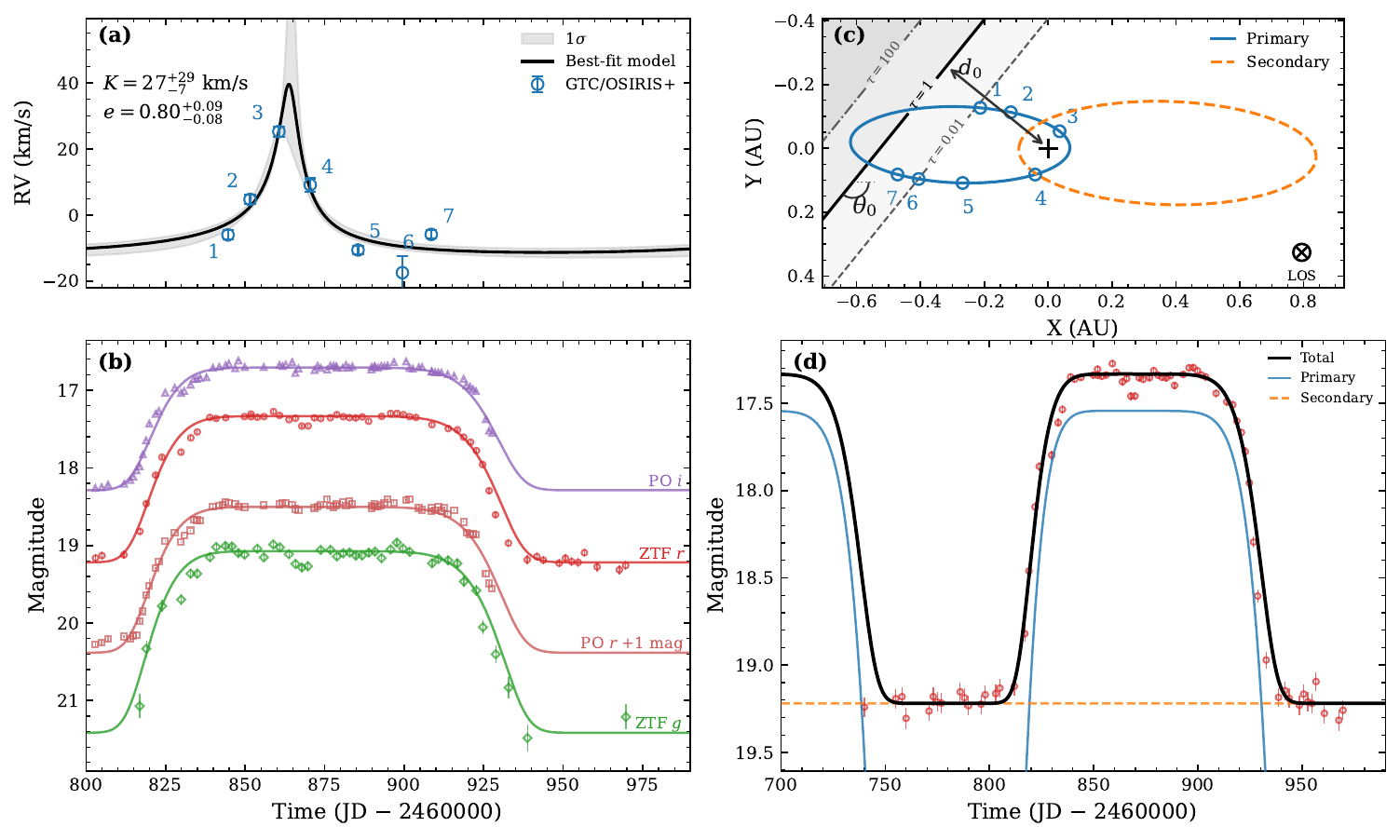}
\caption{Radial velocity and photometric data, best-fit models, and the geometric configuration of Bernhard-1. \textbf{(a)} Radial velocity curve. The black line and gray shading show the best-fit Keplerian model and its $1\sigma$ uncertainty. Blue circles are the GTC/OSIRIS measurements used in the fit. Numbers label the epoch indices. \textbf{(b)} Multi-band light curves. Green diamonds, red circles, pink squares, and purple triangles denote ZTF $g$, ZTF $r$, PO $r$ with $+1$\,mag offset, and PO $i$, respectively. Solid curves are the best-fit semi-transparent screen model in each band. \textbf{(c)} Projected orbital geometry and occultation screen. Blue solid and orange dashed ellipses trace the primary and secondary orbits; the plus sign marks the barycenter. Numbered markers correspond to the RV epochs in panel (a), with the clockwise sense indicating the direction of orbital motion. Note that the $Y$-axis is inverted so that the $Z$-axis points away from the observer. The solid, dashed, and dash-dotted black lines mark the loci where the screen optical depth $\tau = 1$, 0.01, and 100, respectively. The double arrow and arc indicate the projected screen offset $d_0$ and orientation angle $\theta_0$. \textbf{(d)} Decomposed flux contributions in ZTF $r$. The black, blue, and orange dashed curves show the total, primary, and secondary flux, respectively. The primary dominates the occultation depth, while the secondary remains nearly unobscured throughout the orbit.}
\label{fig:rv-lc-schematic}
\end{figure*}

\section{RV modeling} \label{sec:model-rv}

We fit a Keplerian orbit to the seven GTC radial velocities to constrain the orbital elements of the binary. The MMT/MMIRS RV is excluded from the fit because the observation was taken during partial occultation, where the asymmetric obscuration of the stellar disk biases the line profile in a manner analogous to the Rossiter--McLaughlin effect \citep{Rossiter1924,McLaughlin1924, Winn06_orbit_occultations}.

The RV model is parameterized by six quantities: the orbital period $P$, the RV semi-amplitude $K_1$, the eccentricity $e$, the argument of periastron $\omega_1$, the time of periastron passage $T_P$, and the systemic velocity $\gamma$. We fix $P = 191.41$ days, obtained by refitting the \citet{Zhu2022_Two} sharp-edge occultation model to the combined archival and new photometry. The remaining five parameters are sampled with uniform priors over physically permitted ranges using the \texttt{radvel} package \citep{Fulton18_radvel_radial} with \texttt{emcee} \citep{Foreman-Mackey2013_emcee}.

We adopt a fixed RV jitter of 1\,km\,s$^{-1}$ added in quadrature to every measurement uncertainty, as a noise floor for low-resolution spectroscopy. The sixth GTC epoch, which has a larger wavelength calibration uncertainty of $\sim$0.36\,\AA\ , receives conservative larger noise of 5\,km\,s$^{-1}$ to account for potential systematics. We verified this noise treatment by repeating the fit with the jitter as a free parameter drawn from a log-uniform prior. The jitter posterior piles up toward zero, suggesting that the measurement uncertainties introduced by the GP covariance matrix already account for most of the systematics. We nonetheless adopted the 1\,km\,s$^{-1}$ floor to be conservative.

The best-fit orbit is highly eccentric with $e \approx 0.80$. All fitted parameters and uncertainties are listed in Table~\ref{tab:star-prop}. Combining the radial velocity with the SED-derived masses $M_1$ and $M_2$ yields the semi-major axis $a \approx 0.79$\,AU and an orbital inclination $i_{\rm b} \approx 54^\circ$. These quantities, together with $e$, $\omega_1$, and $T_P$, fully specify the projected orbit \textit{shape} on the sky. The orientation of the orbit is still unknown without a measurement of the longitude of the ascending node $\Omega$.

\begin{figure*}[ht!]
\centering
\includegraphics[width=\textwidth]{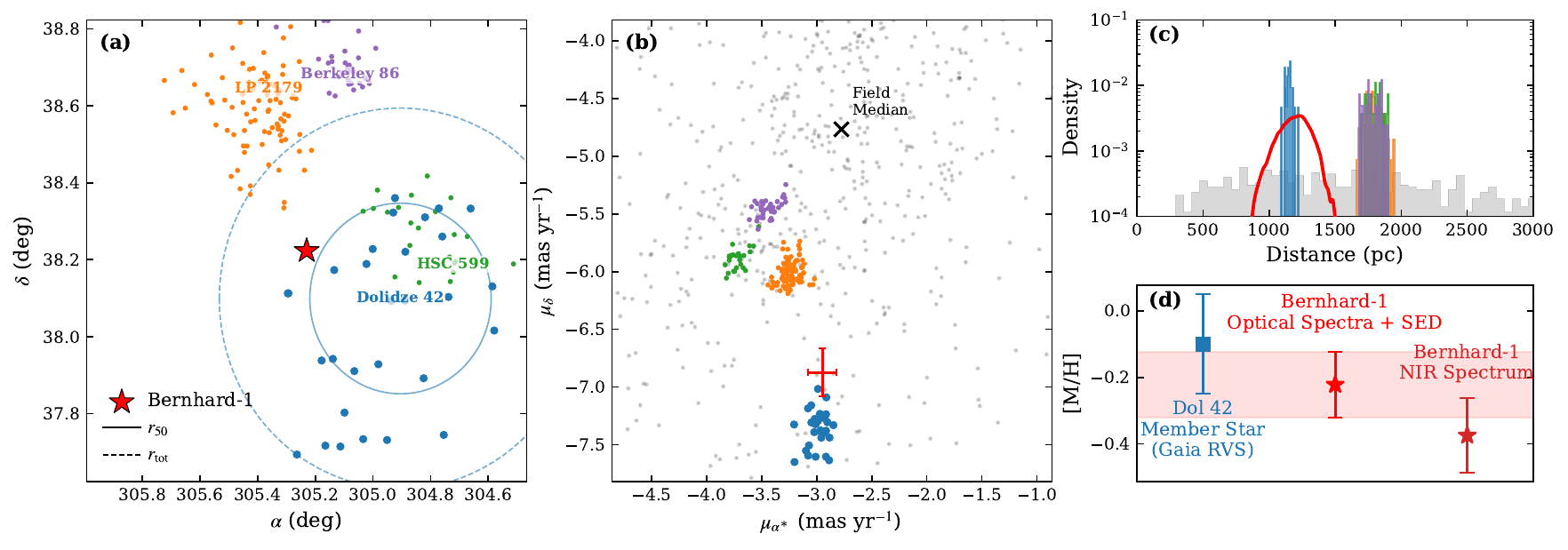}
\caption{The astrometric properties of Bernhard-1 and nearby clusters, together with the comparison of metallicity between Bernhard-1 and Dolidze 42. Panel a: The sky distribution of four closest open clusters, where the solid lines represent the 50\% cluster membership and dashed lines the total cluster radius \citep{Hunt24_improving_open}. Panel b: Proper motions of the identified cluster members, Bernhard-1, and a sample of nearby field stars in Gaia DR3. The field stars are shown as gray points and the median of them are shown as the black cross. Panel c: Normalized histograms of the distance of nearby open clusters and the posterior distribution of the distance constrained by the parallax and the SED. Panel d: The metallicity of one of the Dolidze 42 members and Bernhard-1. Note that Gaia RVS measurements are not available for the members in other nearby open clusters.}
\label{fig:cluster}
\end{figure*}

\section{Light Curve Modeling} \label{sec:model-lc}

We model the multi-band light curves in two steps: the sharp-edge occultation model of \citet{Zhu2022_Two} is used only to refine the orbital period, while a new, physically motivated semi-transparent screen model is used to characterize the occulting material and geometry. The physical model is required because, although the \citet{Zhu2022_Two} model fits the light curve well, it implies an unrealistically slow transverse velocity of $\sim 0.1\,R_\star$ per day, an order of magnitude lower than the apocentric velocity of an equal-mass solar binary with $e = 0.8$ on a 200-day orbit. We also tested the \citet{Winn06_orbit_occultations} model, but this model converges effectively to a similar semi-transparent screen model, as explained at the end of this section.

We first refit the sharp-edge model of \citet{Zhu2022_Two} to the combined archival and new photometry to determine the orbital period. This model assumes an opaque, straight occulting edge and no limb darkening across the stellar photosphere, so that the occultation with orbital period $P$ is fully characterized by the velocities ($v_{\rm in}$, $v_{\rm out}$) and timings ($t_{\rm in}$, $t_{\rm out}$) of the primary star relative to the edge at ingress and egress. We refer the reader to \citet{Zhu2022_Two} for details. The model yields $P = 191.41 \pm 0.07$ days, which is held fixed in the RV modeling of Section~\ref{sec:model-rv}.

Our physical model represents the occulter as a static, semi-transparent screen projected onto the sky plane. We assume the projected column density varies smoothly across the edge relative to the stellar radius. We approximate the variation in optical depth of the screen edge as an exponential profile
\begin{equation}
\tau(s) = \tau_0 \exp\left(-{s / s_0}\right) .
\end{equation}
Here $s_0$ is the characteristic scale length. The exponential profile in optical depth comes naturally out of an exponential projected column-density profile.
We adopt a sky-plane coordinate system centered on the binary center of mass, with the $x$-axis pointing toward the ascending node. The screen edge is specified by its perpendicular offset $d_0$ from the origin and by its orientation angle $\theta_0$, measured from the $x$-axis in the direction of the projected orbital motion of the primary (see panel c of Figure~\ref{fig:rv-lc-schematic}). The edge position $d_0$ is so defined that the optical depth $\tau_0=1$. For an arbitrary sky-plane position $\mathbf r=(x,y)$, we define the signed distance from the edge as
\begin{equation}
s \equiv \hat{\mathbf q}_{\theta}\cdot \mathbf r - d_0
= x \sin \theta_0 - y \cos \theta_0 - d_0,
\end{equation}
where $\hat{\mathbf q}_{\theta} = (\sin\theta_0,\,-\cos\theta_0)$ is the unit normal to the screen edge.
Points with $s>0$ lie on the more transparent side of the screen, where $\tau<1$ and approaches zero far from the edge, whereas points with $s<0$ lie deeper inside the screen, where $\tau>1$.

We treat both stars as point sources, as the scale length $s_0 \approx 0.033\,{\rm AU}$ is several times larger than the stellar radii. The total observed
flux is
\begin{equation}
F_{\rm tot} = F_1\, e^{-\tau(s_1)} + F_2\, e^{-\tau(s_2)},
\end{equation}
where $s_j$ is the signed distance of star $j$ from the screen edge, and $F_1$ and $F_2$ are the unobscured fluxes of the two stars in a given band. In practice, $F_1$ and $F_2$ are determined by linear regression for each trial set of screen parameters, so they are profiled out rather than sampled. We fix the orbital motion of the binary to the best-fit solution from the RV modeling.

Fitting the physical model to the multi-band light curves yields a tightly constrained screen geometry. We fit the ZTF $g$, $r$, and the PO $r$ and $i$ photometry from JD 2460775 to 2460975 simultaneously, sampling the posterior of $(\theta_0, d_0, s_0)$ with \texttt{emcee} \citep{Foreman-Mackey2013_emcee} under uniform priors. The best-fit parameters are listed in Table~\ref{tab:star-prop}, and the best-fit model is compared with the data in panel (b) of Figure~\ref{fig:rv-lc-schematic}. As shown by the flux decomposition in panel (d), the occultation depth is dominated by the primary, while the secondary remains nearly unobscured throughout the orbit.

For completeness, we also apply the model of \citet{Winn06_orbit_occultations}, and conclude that this model does not yield a physically meaningful solution in the case of Bernhard-1. This model describes the occultation of limb-darkened stars, each surrounded by a halo with an exponential brightness profile, by an opaque sharp edge. Applying it to Bernhard-1, the model always converged to an unrealistic solution in which the halo is far brighter than the central stars, and forcing the halo to be fainter than the star would substantially degrade the goodness of the fit. Because the preferred solution obscures an effectively exponential brightness profile with a sharp edge, it is mathematically equivalent to blocking a point source with an exponential transmission profile. We therefore conclude that the \citet{Winn06_orbit_occultations} model qualitatively converges into the same semi-transparent screen model, although it is not exactly the same as the one we have adopted.

\section{Discussion} \label{sec:discussion}

\begin{figure*}[ht!]
\centering
\includegraphics[width=0.7\textwidth]{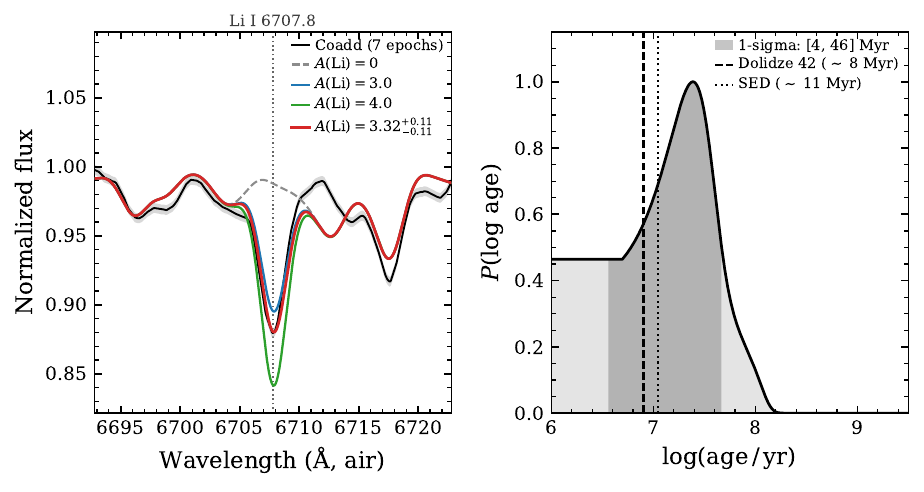}
\caption{The combined GTC spectrum and synthetic spectra around the 6708 \AA Li I line, together with the age determination from the lithium abundance. Left panel: The black solid line represents the combined data, with the gray region indicating the 1$\sigma$ uncertainty. The black dashed line shows the spectrum with no lithium. The spectra in orange, green, and red have lithium abundances A(Li) = 3.0, 3.32 (the best-fit value), and 4.0, respectively. The vertical dashed blue line marks the center of the 6708 \AA Li I line. Right panel: The age determination from lithium abundance using \texttt{eagles} \citep{Jeffries23_gaia-eso_survey}. The posterior is shown by the solid line, with the dark gray shaded region indicating the 1$\sigma$ region. The black dashed and dotted lines indicate the age of Dolidze 42 and the median age in the SED posterior, respectively.}
\label{fig:li}
\end{figure*}

\subsection{Cluster Identification}
\label{subsec:cluster}

To independently constrain the age, metallicity, and distance of Bernhard-1, we investigated whether it is associated with any known open cluster (OC) or moving group (MG). We compared the system with the OC/MG catalog of \citet{Hunt24_improving_open} by cross-matching it using the Gaia DR3 sky coordinates and proper motions, along with the SED-derived distance that employs the Gaia parallax as a prior.

A direct membership check first shows that the Gaia DR3 source ID of Bernhard-1 is not listed as a member of any cataloged OC or MG. We therefore searched for all OC/MGs whose centers lie within 30\arcmin\ of Bernhard-1, yielding six spatial candidates. Of these, only four fall within 10$\sigma$ of Bernhard-1 in proper-motion space, accounting for both the measurement uncertainty and the intrinsic cluster scatter, and only Dolidze 42 (hereafter D42) lies within the 3$\sigma$ region (panel b of Figure~\ref{fig:cluster}). The distance provides an independent filter with the same outcome, that only D42 agrees with Bernhard-1 within 3$\sigma$ (panel c). We use the metallicity as a final sanity check. Gaia RVS metallicities are unavailable for the members of most nearby candidate clusters. Among the candidates considered here, only one D42 member has a Gaia RVS metallicity measurement, and its metallicity is broadly consistent with our GTC and MMT values (panel d).

Despite the qualitative agreement between properties of D42 and Bernhard-1, the membership probability remains limited by the distance uncertainty. To quantify it, we trained a seven-component Gaussian mixture model (GMM) in proper-motion-parallax space, fixing six components to the means and covariances of the member astrometry of each candidate cluster and fitting one additional component to the field-star population. Drawing $10^4$ samples from the Bernhard-1 posterior in this space yields a $\sim$ 30\% probability that Bernhard-1 is a D42 member, against $\sim $ 70\% for the field. This modest probability is driven almost entirely by the distance uncertainty, which is large compared with the distance scatter of the D42 members. As shown in the following subsection, the lithium- and SED-derived ages are nonetheless consistent with the age of D42, which independently supports the association.

\subsection{Lithium Abundance and Age Determination}

The GTC/OSIRIS spectra cover the Li I 6708~\AA\ line, which provides an independent age estimate that tests both the youth of Bernhard-1 and its association with D42. Following the procedure adopted for Bernhard-2 \citep{Hu24_eccentric_binary}, we measured the lithium abundance using the same spectral-synthesis settings as in Section~\ref{subsec:spec-fit}. We built a grid spanning $A({\rm Li}) = 0.0$--$6.0$ in steps of 0.1~dex and effective resolving power $R = 1500$--$3500$ in steps of 100, over the 668.6--672.6~nm range, with all other stellar parameters fixed to the values in Table~\ref{tab:star-prop}. During the fit, we interpolate within this grid, multiply the model by a third-order polynomial to account for the local continuum, and include a jitter term to absorb residual systematics. This yields a lithium abundance of $A({\rm Li}) = 3.32 \pm 0.11$, as shown in the left panel of Figure~\ref{fig:li}.

The lithium abundance implies a young age consistent with the pre-main-sequence interpretation of Bernhard-1. We convert the best-fit lithium feature into an equivalent width by integrating the flux difference between the best-fit synthetic spectrum and a zero-lithium model, and then infer the age using \texttt{eagles} \citep{Jeffries23_gaia-eso_survey}. The resulting 1$\sigma$ age range is 4--46 Myr, as shown in the right panel of Figure~\ref{fig:li}. This lithium age is consistent with both the SED-derived age ($\sim 11$ Myr) and the age of D42 ($\sim 8$ Myr), reinforcing the cluster association inferred in Section \ref{subsec:cluster} and confirming the pre-main-sequence nature of the system.

\subsection{Mutual Inclination Between the Disk and Binary}
\label{subsec:mut-inc}

\begin{figure}[ht!]
\centering
\includegraphics[width=0.47\textwidth]{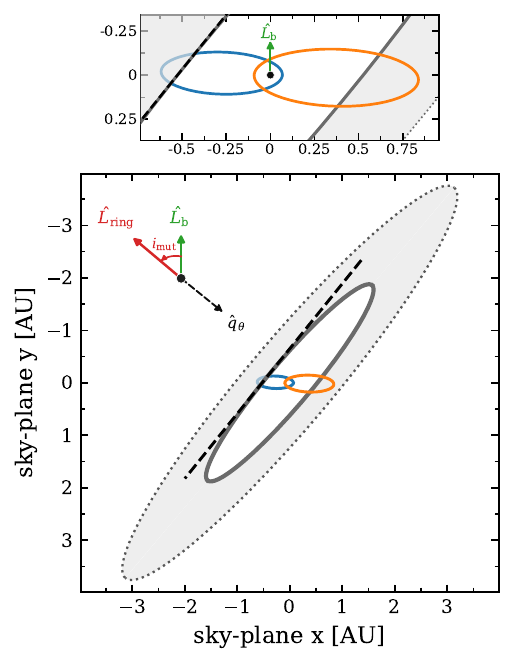}
\caption{
Schematic geometry of the binary orbit and a misaligned circumbinary disk with $i_{\rm mut}=52^\circ$. The lower panel shows the sky-plane projection of the binary orbits and the circumbinary disk, with the primary and secondary shown in blue and orange, respectively. The black dashed line marks the best-fit occultation edge from the semi-transparent screen model, while the solid gray ellipse shows a circular disk ring whose projected tangent matches this edge. The dotted gray ellipse illustrates another possible disk edge, whose location is unconstrained and is shown only for visualization. The occultation-edge normal $\hat{q}_{\theta}$ and the angular-momentum unit vectors of the ring and binary are shown in the upper-left corner. The upper panel zooms in on the binary and occultation edge, illustrating that the projected ring is locally well approximated by a straight screen. The binary rotates clockwise, using the same sky-plane convention as in Figure~\ref{fig:rv-lc-schematic}.
}
\label{fig:geo}
\end{figure}

The phenomenological screen parameters ($d_0$ and $\theta_0$) derived from the light curve provide a geometric constraint on the disk, regardless of which specific part of the disk causes the occultation. Misaligned circumbinary disks typically have low eccentricities ($e \lesssim 0.1$) that decrease further from the central binary \citep{Smallwood22_accretion_binary, Ragusa20_evolution_large}. Therefore, a warped circumbinary disk can be conceptualized as a series of concentric circular rings, only with different inclinations and twists. The occulting edge, which could be the inner edge, outer edge, or even a warped structure, still corresponds to one of these rings. Because our goal is to provide a rough estimate of the overall disk--binary mutual inclination, treating the occulting structure as a representative circular ring is physically well-justified. Furthermore, because the projected distance from the binary barycenter to the screen edge $d_0$ is much smaller than the expected disk inner edge truncation radius \citep[$\gtrsim 2 a_{\rm b}$, ][]{Miranda2015_tidal_truncation}, the local curvature of the disk edge is negligible. The screen can therefore be safely modeled as a straight line tangent to this projected representative ring, as illustrated in Figure~\ref{fig:geo}.

The edge offset alone requires the occulting ring to lie close to edge-on and gives a conservative lower limit on the mutual inclination. A circular ring of radius $R_{\rm ring}$ and sky-plane inclination $i_{\rm ring}$ projects to an ellipse with semiminor axis $R_{\rm ring}|\cos i_{\rm ring}|$. The ring inclination is therefore bounded by
$i_{\rm ring} \geq \arccos\left({|d_0|}/{R_{\rm ring}}\right)$. For the fiducial choice $R_{\rm ring}=3a_{\rm b}\simeq2.4$~AU and the measured $|d_0|\simeq0.4$~AU, this gives $i_{\rm ring}\gtrsim80^\circ$. Combining this result with $i_{\rm b}\simeq54^\circ$ yields
$i_{\rm mut} \geq | i_{\rm ring}-i_{\rm b}| \gtrsim26^\circ$, independent of the longitude of the ascending node of the occultation ring. Adopting $R_{\rm ring}=2$--$3a_{\rm b}$ gives a similar lower limit of approximately $21^\circ$--$26^\circ$.

The measured edge orientation provides a substantially stronger constraint than the offset alone. We define the unit vector normal to the occultation edge in the sky plane as $\hat{q}_{\theta} = (\sin\theta_0,-\cos\theta_0,0)$. Because the $x$-axis is defined along the ascending node of the binary orbit, the binary angular-momentum unit vector is $\hat{L}_{\rm b} = (0,-\sin i_{\rm b},\cos i_{\rm b})$. In the exactly edge-on limit, the angular momentum of the occulting ring lies in the sky plane and is normal to the projected ring edge, such that $\hat{L}_{\rm ring} = \pm\hat{q}_{\theta}.$ The mutual inclination then satisfies 
\begin{equation} 
\cos i_{\rm mut} = 
\hat{L}_{\rm ring} \cdot \hat{L}_{\rm b} = \pm\sin i_{\rm b}\cos\theta_0. 
\end{equation} 
Using the measured values of $i_{\rm b}$ and $\theta_0$ gives $i_{\rm mut}\simeq120^\circ$ or $180^\circ-i_{\rm mut}\simeq60^\circ$. These two solutions reflect the fact that the projected occultation geometry constrains the ring plane but cannot determine the sign of its angular-momentum vector, and therefore cannot distinguish between the corresponding prograde-like and retrograde-like configurations. The finite screen offset changes the edge-on estimate by only a modest amount as $|90^\circ - i_{\rm ring}| \ll i_{\rm mut}$ in the nearly edge-on case. Applying the $d_0$ correction with a fiducial $R_{\rm ring}=3a_{\rm b}$ yields a mutual inclination of $i_{\rm mut} \sim 50^\circ$ or $130^\circ$, and the case with $i_{\rm mut} = 52^\circ$ is shown in Figure~\ref{fig:geo}.

All geometries consistent with the occultation edge therefore identify Bernhard-1 as a highly misaligned circumbinary-disk system. Its inferred mutual inclination is substantially larger than the $\sim10^\circ$ misalignment measured for KH~15D \citep{Winn06_orbit_occultations, Poon21_constraining_circumbinary}. Such a large tilt, together with the high eccentricity of the central binary, places Bernhard-1 in the dynamical regime where evolution toward a polar configuration may be possible \citep[e.g.,][]{Martin17_polar_alignment, Zanazzi18_inclination_evolution, Childs21_formation_polar}. More generally, the same approach can be applied to CBO systems for which RV monitoring yields a binary orbit and photometric modeling constrains an occulting edge. A larger sample of RV-confirmed CBO systems would therefore provide a direct way to measure the mutual-inclination distribution of young circumbinary disks.

\subsection{Disk Precession and Occultation Color Change From the Light Curve}

\begin{figure*}[ht!]
\centering
\includegraphics[width=\textwidth]{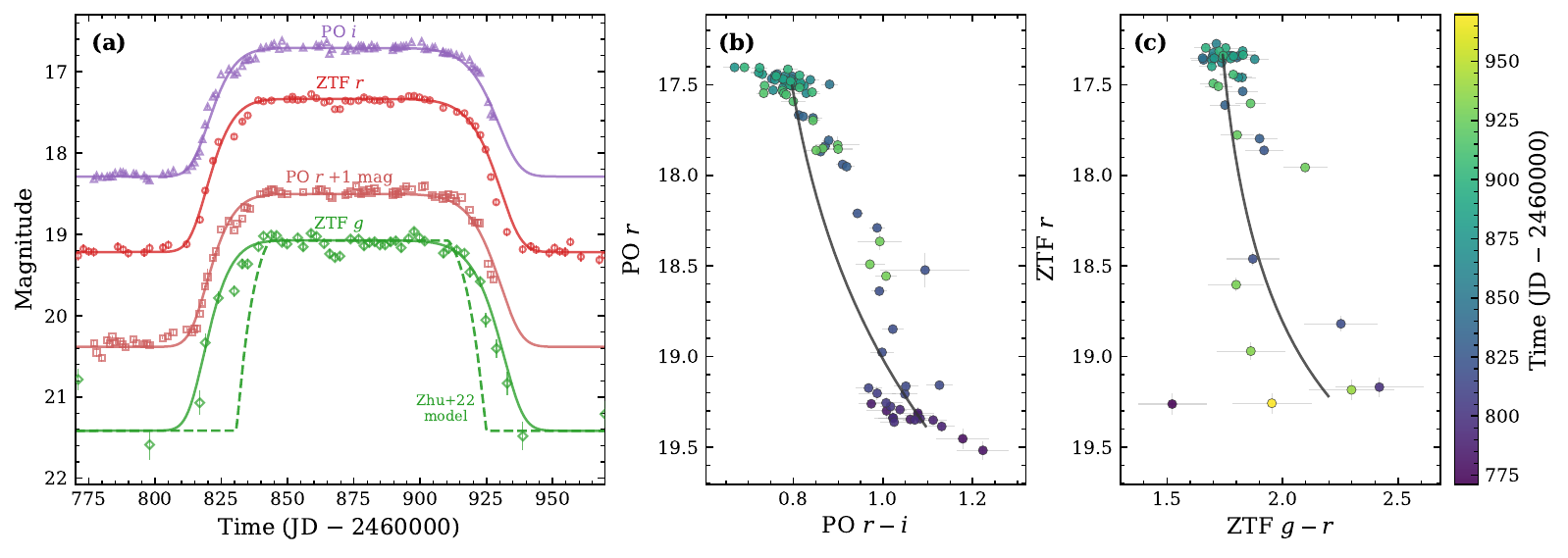}
\caption{The light curve with the new model and \citet{Zhu2022_Two} model, together with the corresponding color magnitude variation for the occultation process in panels (b) and (c). The label in panel (a) is the same as in panel (b) of Figure \ref{fig:rv-lc-schematic}, except for the green dashed line, which represents the \citet{Zhu2022_Two} model in the ZTF g band and shows a longer occultation duration. Panels (b) and (c): CMD of the PO $r$ and $i$ observations and the ZTF g and r observations. The color indicates the time of the observation. CMD points are plotted only when both bands were obtained on the same night. The black solid line in the two panels indicates the model CMD track corresponding to the solid line in panel (a).}
\label{fig:lc-cmd}
\end{figure*}

The multi-band light curves show both long-term occultation evolution and weak optical chromaticity. In the following, we first quantify the timing evolution, and then use the color--magnitude behavior to test the gray-occultation approximation and discuss possible dust constraints.

The new occultation is shorter than that observed around 2020, likely due to a precessing misaligned disk. We compare our semi-transparent screen model, fit to the multi-band photometry from JD~2460775 to 2460975, with the best-fit \citet{Zhu2022_Two} model in panel (a) of Figure~\ref{fig:lc-cmd}. The occultation duration is now clearly shorter than in the earlier epoch. The same gradual drift is seen in Bernhard-2 \citep{Hu24_eccentric_binary} and KH~15D \citep[e.g., ][]{Winn06_orbit_occultations, GarciaSoto2020, Poon21_constraining_circumbinary}, where it is naturally explained by the precession of a misaligned circumbinary disk.

\begin{figure*}[ht!]
\centering
\includegraphics[width=0.9\textwidth]{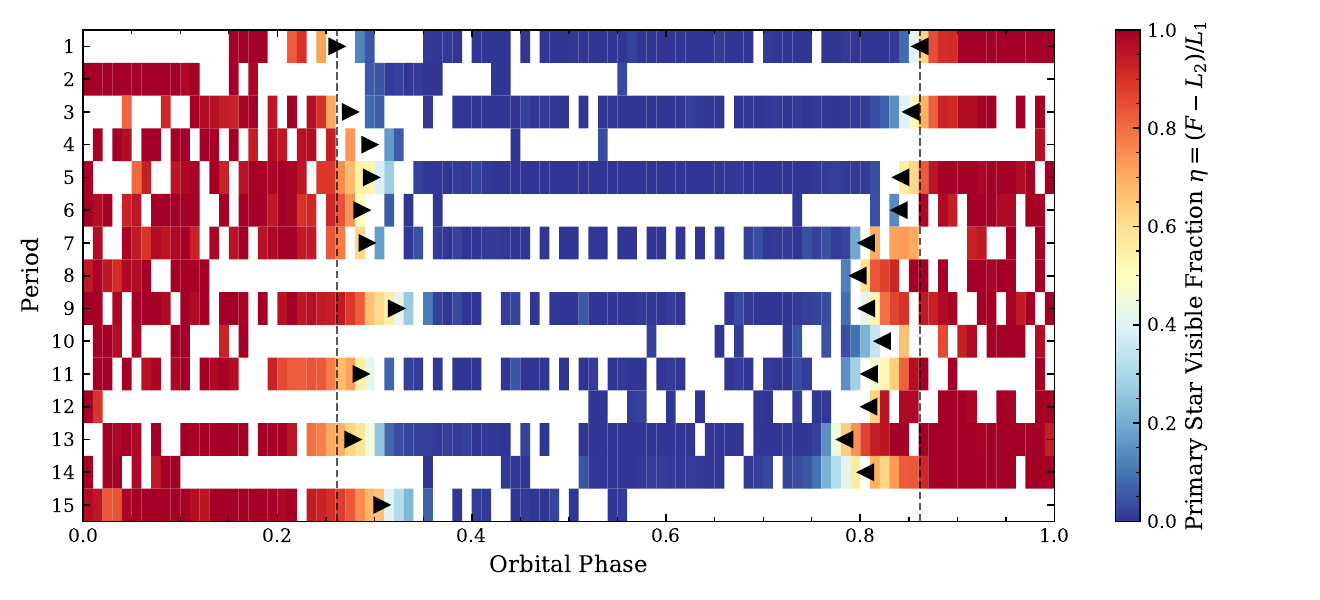}
\caption{Evolution of the occultation morphology across different orbital cycles, shown as a river plot. The color represents the visible fraction of the primary-star flux, with blue corresponding to nearly complete occultation and red corresponding to an unobscured primary. Rightward and leftward triangles mark the best-fit ingress and egress timings of individual cycles, respectively. The error bars are smaller than the markers. The two vertical dashed lines mark the ingress and egress timings of the first observed ZTF cycle.}
\label{fig:river}
\end{figure*}

We construct a ``river plot'' to compare the observed data and modeled timings. We quantify the occultation timing evolution by converting the optical light curves into the visible fraction of the primary star. For each band, we define \begin{equation} \eta(t) = {F(t) - F_2 \over F_1}, \end{equation} where $F(t)$ is the instantaneous observed flux in that band. This normalization allows us to combine the ZTF and PO observations into a common light-curve diagnostic under the approximation that the optical occultation is nearly achromatic, which will be justified in the color discussion below. We divide each orbital cycle into 100 phase bins, corresponding to a bin width of about two days, and assign each bin the median value of $\eta$ from all valid observations in that bin. A color is assigned for each bin from blue (low $\eta$) to red (high $\eta$). We then fit the \citet{Zhu2022_Two} sharp-edge model separately to each cycle, with the component fluxes fixed, to estimate the ingress and egress timings whenever the corresponding phase range is sufficiently sampled. These are marked by rightward and leftward triangles in Figure~\ref{fig:river}.

The occultation duration shows two behaviors: an overall shrinking trend superposed with non-monotonic variations on a few-period timescale. The black dashed lines in Figure~\ref{fig:river} mark the ingress and egress timings of the first observed ZTF period. All later durations fall within this range and tend to shorten in recent periods. Yet the trend is not monotonic: the duration in cycle 11 ($\sim$101 days) exceeds that in cycle 9 ($\sim$93 days). A Lomb-Scargle analysis of the ingress and egress timings, after removing the orbital-period alias, shows a tentative peak near 1000 days. The overall shrinking matches the expectation for a precessing disk, whereas the short-timescale variation is far shorter than the disk--binary secular interaction timescale. These deviations could reflect local disk structure, or opacity variations, but the present data are insufficient to assign a physical interpretation to this signal.

Combining the timing change with the RV orbit gives an order-of-magnitude precession rate. The change appears in two geometric signatures. First, the shorter duration corresponds to an increase in the screen offset $d_0$ of about 0.1~AU, obtained by fitting the screen model to the earlier ZTF data. For an occultation edge at $\sim 3a \approx 2$~AU, this implies an angular change of about $3^\circ$. Second, the egress shifts by about 15~days while the ingress shifts by less than 10~days, corresponding to a decrease in the screen angle $\theta_0$ of about $10^\circ$, which is larger than the offset-based estimation. Since the combined data span roughly 15 orbits, the precession rate implied by $\theta_0$ alone is about $0.7^\circ$ per binary orbit, or $\sim 1^\circ$~yr$^{-1}$. This is of the same order as the precession observed in KH~15D \citep{Winn06_orbit_occultations,Poon21_constraining_circumbinary}. Continued monitoring is needed to refine these values and to constrain disk parameters such as the scale of the occultation disk.

The occultation is also mildly redder than a gray process. We construct color--magnitude diagrams of the occultation for both the PO and ZTF photometry, shown in panels (b) and (c) of Figure~\ref{fig:lc-cmd}, using only observations obtained on the same night so as to remove the color change induced by orbital motion between nights. The resulting time separations are less than an hour for PO and less than two hours for ZTF. The black solid line marks the track expected for an achromatic occultation in flux space, and the PO data show tentative evidence for reddening relative to this gray track at ingress/egress.

This chromaticity is small enough to validate the achromatic assumption used in the river plot. The $r-i$ difference is only about 0.1~mag, which translates to about one day in occultation timing, and is much shorter than the multi-period timescale of the timing change itself. Future multi-band monitoring, particularly in the infrared, will be able to infer the extinction law of the occultation material and thereby constrain the dust grain size, as has been done for KH~15D \citep[e.g., ][]{Arulanantham16_seeing_through, GarciaSoto2020}.

\subsection{Pulsed Accretion from H$\alpha$}

\begin{figure*}[ht!]
\centering
\includegraphics[width=0.85\textwidth]{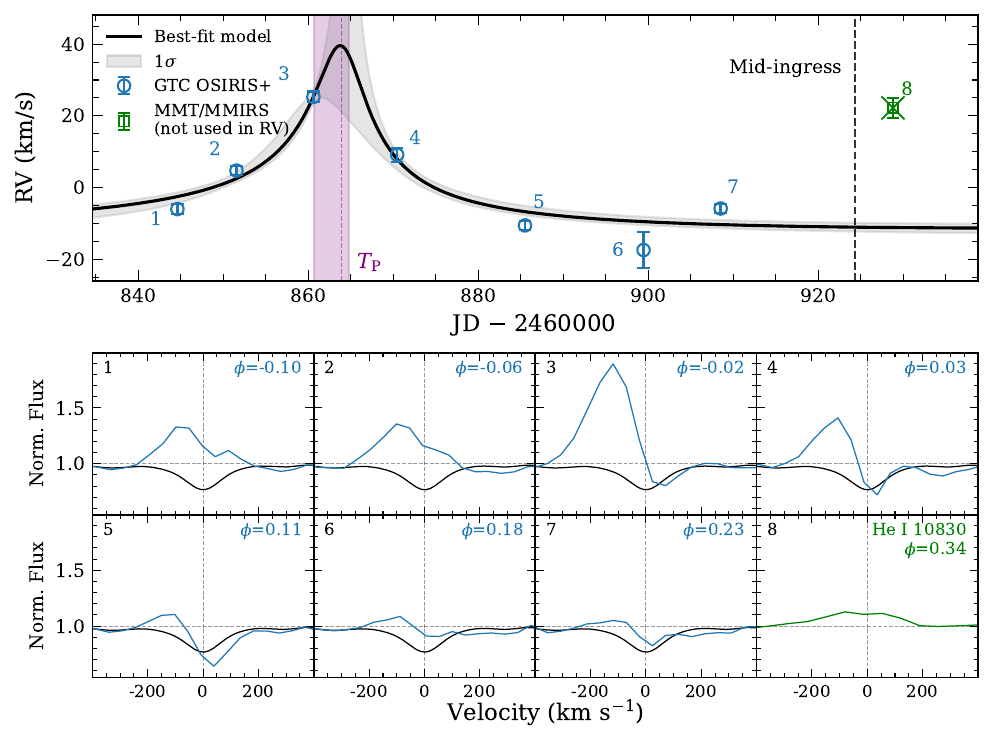}
\caption{The RV fitting result (upper panel) and the corresponding spectral profiles near the accretion indicators (lower panels). The description for the data and model in the upper panel is the same as in Figure \ref{fig:rv-lc-schematic}. In the lower panels, the blue solid lines and the green solid line represent the observations from GTC and MMT, respectively, while the black solid line corresponds to the template of the best-fit primary star. The horizontal and vertical black dashed lines indicate the continuum and the line center, respectively. The numbers in the upper left corner indicate the observation index, and the numbers in the upper right corner indicate the phase of the observation.}
\label{fig:halpha}
\end{figure*}

The H$\alpha$ line profiles observed by GTC/OSIRIS vary with orbital phase and show enhanced, inverse P-Cygni profile near pericenter, indicating pulsed accretion. This behavior closely resembles that of Bernhard-2 \citep{Hu24_eccentric_binary} and of nearly aligned or coplanar circumbinary systems such as KH~15D and DQ~Tau \citep[][]{Hamilton12_complex_variability, Tofflemire17_accretion_magnetic}. The corresponding line profiles are shown in Figure~\ref{fig:halpha}.

The H$\alpha$ morphology evolves from emission-dominated profiles after egress to inverse P-Cygni profiles closer to periastron. Epochs 1 and 2, obtained about 20 days after egress, show mainly emission-only profiles. In the later epochs, the line develops a redshifted absorption component, with the most prominent accretion-like profile appearing at epoch 3, close to periastron passage. Using the empirical relation between the H$\alpha$ 10\% width, WH$\alpha$(10\%), and the mass-accretion rate, we estimate an accretion scale of order $10^{-10}\,M_\sun\,{\rm yr}^{-1}$ for the first three epochs \citep{Alcala14_x-shooter_spectroscopy}. Because the line profiles are complex and our spectra have moderate resolution, we use this relation only as an order-of-magnitude accretion diagnostic.

The timing of this accretion enhancement is consistent with expectations for an eccentric, misaligned circumbinary disk. In the coplanar case, pulsed accretion is understood as a two-step process: the accretion stream from the circumbinary disk is enhanced near apocenter, while near pericenter the strong tidal and magnetospheric interactions between the stars promote angular-momentum loss and rapid magnetospheric accretion \citep[e.g., ][]{Artymowicz96_mass_flow, Basri97_classical_t, Munoz16_pulsed_accretion, Tofflemire17_accretion_magnetic}. In the misaligned case, instead, the accretion stream from the disk is enhanced near the nodes of the binary orbit relative to the disk plane \citep[e.g., ][]{Smallwood23_formation_polar}, and the accretion onto the star is similar to that in the coplanar case. For the highly eccentric orbit of Bernhard-1, the node lies close to pericenter, and the two effects seem to coincide and produce enhanced accretion generally near pericenter.

Our MMT/MMIRS observation taken during the ingress of the occultation covers both the He~I 10830 line and several Paschen lines, all of which could serve as tracers of accretion \citep[e.g., ][]{Alcala14_x-shooter_spectroscopy}. We find that the He line shows a broad emission feature with a full width at half maximum (FWHM) of about 300~km~s$^{-1}$, whereas the Paschen lines appear in absorption and are consistent with the stellar template. Therefore, the broad, emission-only He profile is more likely a signature of an accretion-powered stellar wind or jet rather than of accretion itself \citep{Edwards06_probing_t, Erkal22_he_i, Kwan2007_modeling_t}.

\section{Conclusion}
\label{sec:conclusion}

Our new spectroscopic and multi-band photometric observations confirm Bernhard-1 as a CBO system, the third after KH~15D and Bernhard-2 \citep{Johnson2004_KH15D_binary, Winn06_orbit_occultations, Hu24_eccentric_binary}. The radial velocities reveal a highly eccentric binary ($e = 0.80 \pm 0.09$) composed of two pre-main-sequence K dwarfs of $\sim 1.1\,M_\odot$ and $\sim 0.8\,M_\odot$, whose periodic dimming arises from occultation by a misaligned circumbinary disk. The lithium and SED ages both indicate a young system of $\sim$10~Myr. Furthermore, its spatial, astrometric, and metallicity properties are all consistent with the open cluster Dolidze~42, suggesting that Bernhard-1 could be a member of this $\sim8$~Myr cluster.

Combining the RV orbit with the occultation-screen geometry yields a rough disk--binary mutual inclination of $i_{\rm mut} \sim 50^\circ$ or $130^\circ$, with the degeneracy arising from the unknown disk rotation direction. This mutual inclination is substantially larger than the $\sim10^\circ$ measured for KH 15D, the well-studied prototype of CBO systems. Considering its high eccentricity, Bernhard-1 likely will evolve into a polar aligned configuration. This geometric method can be applied to any CBO system once RV monitoring delivers an orbital solution, opening a route to measuring the mutual-inclination distribution of young circumbinary disks.

Like KH~15D and Bernhard-2, Bernhard-1 displays both ongoing disk precession and pulsed accretion near pericenter. The light curve shows a shorter occultation than was observed around 2020, implying a precession rate of about 0.7$^\circ$ per binary orbital period, while the phase-dependent H$\alpha$ profiles show enhanced inverse P-Cygni profile near pericenter. All three spectroscopically confirmed CBO systems thus share these two features. The precession is expected, although the differences in rate among systems require dedicated modeling. The persistence of pulsed accretion, on the other hand, may indicate that the general accretion pattern observed in coplanar systems also applies to misaligned circumbinary disks.

Bernhard-1 nonetheless joins a still small, statistically limited, and yet scientifically valuable sample: to date, only three of more than 30 proposed CBO candidates have been confirmed via spectroscopy. Systematic RV monitoring is required to increase the size of this sample. Such observations will provide constraints on the distribution of primordial disk–binary alignments and, in turn, on the initial conditions governing the formation of circumbinary planets.

\begin{acknowledgments}

We thank Jeremy Smallwood, Ruiqi Yang, J.J. Zanazzi, and Gabriella Zsidi for useful discussions. Z.H., W. Zhu and W. Zang acknowledge support by the National Natural Science Foundation of China (grant No. 12133005). P.C. acknowledges support from the Zhejiang Provincial Top-Level Research Support Program. Observations reported here were obtained at the MMT Observatory, a joint facility of the Smithsonian Institution and the University of Arizona. This paper uses data products produced by the OIR Telescope Data Center, supported by the Smithsonian Astrophysical Observatory. This research benefited from Z.H.'s participation in the 2026 workshop ``Exploring Planetary Systems in the Era of Time-domain Astronomy'', hosted by the Institute for Astronomy at the University of Hawai'i and supported by award 2106927 from the National Science Foundation's Astronomy \& Astrophysics Research Grants program.

\end{acknowledgments}

\vspace{5mm}
\facilities{GTC (OSIRIS), MMT (MMIRS), ZTF, \gaia, 2MASS, Pan-STARRS, WISE}
\software{\texttt{astropy} \citep{astropy:2013,astropy:2018,astropy:2022}, 
          \texttt{scipy} \citep{Virtanen2020_SciPy}, 
          \texttt{emcee} \citep{Foreman-Mackey2013_emcee}, 
          \texttt{isochrones} \citep{Morton2015_isochrones},
          \texttt{PypeIt} \citep{pypeit:joss_pub, pypeit:zenodo}
          \texttt{fitramp} \citep{Brandt24_optimal_fitting, Brandt24_likelihood-based_jump}
          \texttt{iSpec} \citep{Blanco-Cuaresma2014_iSpec, Blanco-Cuaresma2019_iSpec2},
          \texttt{starfish} \citep{Czekala2015_Constructing},
          DoPHOT \citep{Schechter1993_DoPHOT, AlonsoGarcia2012_DoPHOT_C},
          \texttt{radvel} \citep{Fulton18_radvel_radial},
          \texttt{eagles} \citep{Jeffries23_gaia-eso_survey}
          }

\bibliography{main}{}

@INPROCEEDINGS{Cepa2010_GTC_OSIRIS,
       author = {{Cepa}, Jordi},
        title = "{OSIRIS: Final Characterization and Scientific Capabilities}",
    booktitle = {Highlights of Spanish Astrophysics V},
         year = 2010,
       editor = {{Diego}, Jose M. and {Goicoechea}, Luis J. and {Gonz{\'a}lez-Serrano}, J. Ignacio and {Gorgas}, Javier},
       series = {Astrophysics and Space Science Proceedings},
       volume = {14},
        month = jan,
        pages = {15},
          doi = {10.1007/978-3-642-11250-8_2},
       adsurl = {https://ui.adsabs.harvard.edu/abs/2010ASSP...14...15C}
}

@article{pypeit:joss_pub,
  doi       = {10.21105/joss.02308},
  url       = {https://doi.org/10.21105/joss.02308},
  year      = {2020},
  publisher = {The Open Journal},
  volume    = {5},
  number    = {56},
  pages     = {2308},
  author    = {{Prochaska}, J. Xavier  and Joseph F. Hennawi and Kyle B. Westfall and Ryan J. Cooke and Feige Wang and Tiffany Hsyu and Frederick B. Davies and Emanuele Paolo Farina and Debora Pelliccia},
  title     = {PypeIt: The Python Spectroscopic Data Reduction Pipeline},
  journal   = {Journal of Open Source Software}
}

@misc{pypeit:zenodo,
  author    = {{Prochaska}, J. Xavier and {Hennawi}, Joseph and {Cooke}, Ryan and
               {Westfall}, Kyle and {Wang}, Feige and {EmAstro} and {Tiffanyhsyu} and
               {Wasserman}, Asher and {Villaume}, Alexa and {Marijana777} and
               {Schindler}, JT and {Young}, David and {Simha}, Sunil and
               {Wilde}, Matt and {Tejos}, Nicolas and {Isbell}, Jacob and
               {Fl{\"o}rs}, Andreas and {Sandford}, Nathan and {Vasovi{\'c}}, Zlatan and
               {Betts}, Edward and {Holden}, Brad},
  title     = {{pypeit/PypeIt: Release 1.0.0}},
  year      = 2020,
  month     = apr,
  eid       = {10.5281/zenodo.3743493},
  doi       = {10.5281/zenodo.3743493},
  version   = {v1.0.0},
  publisher = {Zenodo},
  adsurl    = {https://ui.adsabs.harvard.edu/abs/2020zndo...3743493P}
}

@article{Hu24_eccentric_binary,
  author        = {{Hu}, Zhecheng and {Zhu}, Wei and {Dai}, Fei and {Chen}, Ping and {Huang}, Yang and {Fang}, Min and {Post}, Richard S.},
  title         = {{An Eccentric Binary with a Misaligned Circumbinary Disk}},
  journal       = {\apjl},
  year          = 2024,
  month         = dec,
  volume        = {977},
  number        = {1},
  eid           = {L28},
  pages         = {L28},
  doi           = {10.3847/2041-8213/ad94e8},
  archiveprefix = {arXiv},
  eprint        = {2409.18296},
  primaryclass  = {astro-ph.SR},
  adsurl        = {https://ui.adsabs.harvard.edu/abs/2024ApJ...977L..28H}
}

@article{Mcleod2012_MMT_MMIRS,
  author        = {{McLeod}, Brian and {Fabricant}, Daniel and {Nystrom}, George and {McCracken}, Ken and {Amato}, Stephen and {Bergner}, Henry and {Brown}, Warren and {Burke}, Michael and {Chilingarian}, Igor and {Conroy}, Maureen and {Curley}, Dylan and {Furesz}, Gabor and {Geary}, John and {Hertz}, Edward and {Holwell}, Justin and {Matthews}, Anne and {Norton}, Tim and {Park}, Sang and {Roll}, John and {Zajac}, Joseph and {Epps}, Harland and {Martini}, Paul},
  title         = {{MMT and Magellan Infrared Spectrograph}},
  journal       = {\pasp},
  year          = 2012,
  month         = dec,
  volume        = {124},
  number        = {922},
  pages         = {1318},
  doi           = {10.1086/669044},
  archiveprefix = {arXiv},
  eprint        = {1211.6174},
  primaryclass  = {astro-ph.IM},
  adsurl        = {https://ui.adsabs.harvard.edu/abs/2012PASP..124.1318M}
}

@article{Chilingarian15_data_reduction,
  author        = {{Chilingarian}, Igor and {Beletsky}, Yuri and {Moran}, Sean and {Brown}, Warren and {McLeod}, Brian and {Fabricant}, Daniel},
  title         = {{Data Reduction Pipeline for the MMT and Magellan Infrared Spectrograph}},
  journal       = {\pasp},
  year          = 2015,
  month         = apr,
  volume        = {127},
  number        = {950},
  pages         = {406},
  doi           = {10.1086/680598},
  archiveprefix = {arXiv},
  eprint        = {1503.07504},
  primaryclass  = {astro-ph.IM},
  adsurl        = {https://ui.adsabs.harvard.edu/abs/2015PASP..127..406C}
}

@article{Brandt24_optimal_fitting,
  author        = {{Brandt}, Timothy D.},
  title         = {{Optimal Fitting and Debiasing for Detectors Read Out Up-the-Ramp}},
  journal       = {\pasp},
  year          = 2024,
  month         = apr,
  volume        = {136},
  number        = {4},
  eid           = {045004},
  pages         = {045004},
  doi           = {10.1088/1538-3873/ad38d9},
  archiveprefix = {arXiv},
  eprint        = {2309.08753},
  primaryclass  = {astro-ph.IM},
  adsurl        = {https://ui.adsabs.harvard.edu/abs/2024PASP..136d5004B}
}

@article{Brandt24_likelihood-based_jump,
  author        = {{Brandt}, Timothy D.},
  title         = {{Likelihood-based Jump Detection and Cosmic Ray Rejection for Detectors Read Out Up-the-ramp}},
  journal       = {\pasp},
  year          = 2024,
  month         = apr,
  volume        = {136},
  number        = {4},
  eid           = {045005},
  pages         = {045005},
  doi           = {10.1088/1538-3873/ad38da},
  archiveprefix = {arXiv},
  eprint        = {2404.01326},
  primaryclass  = {astro-ph.IM},
  adsurl        = {https://ui.adsabs.harvard.edu/abs/2024PASP..136d5005B}
}

@article{Czekala2015_Constructing,
  title         = {Constructing {{A Flexible Likelihood Function For Spectroscopic Inference}}},
  author        = {Czekala, Ian and Andrews, Sean M. and Mandel, Kaisey S. and Hogg, David W. and Green, Gregory M.},
  year          = 2015,
  month         = oct,
  journal       = {The Astrophysical Journal},
  volume        = {812},
  number        = {2},
  eprint        = {1412.5177},
  primaryclass  = {astro-ph},
  pages         = {128},
  issn          = {1538-4357},
  doi           = {10.1088/0004-637X/812/2/128},
  urldate       = {2026-03-13},
  archiveprefix = {arXiv},
  langid        = {english}
}

@article{Blanco-Cuaresma2014_iSpec,
  author        = {{Blanco-Cuaresma}, S. and {Soubiran}, C. and {Heiter}, U. and {Jofr{\'e}}, P.},
  title         = {{Determining stellar atmospheric parameters and chemical abundances of FGK stars with iSpec}},
  journal       = {\aap},
  year          = 2014,
  month         = sep,
  volume        = {569},
  eid           = {A111},
  pages         = {A111},
  doi           = {10.1051/0004-6361/201423945},
  archiveprefix = {arXiv},
  eprint        = {1407.2608},
  primaryclass  = {astro-ph.IM},
  adsurl        = {https://ui.adsabs.harvard.edu/abs/2014A&A...569A.111B}
}

@article{Blanco-Cuaresma2019_iSpec2,
  author        = {{Blanco-Cuaresma}, Sergi},
  title         = {{Modern stellar spectroscopy caveats}},
  journal       = {\mnras},
  year          = 2019,
  month         = jun,
  volume        = {486},
  number        = {2},
  pages         = {2075-2101},
  doi           = {10.1093/mnras/stz549},
  archiveprefix = {arXiv},
  eprint        = {1902.09558},
  primaryclass  = {astro-ph.SR},
  adsurl        = {https://ui.adsabs.harvard.edu/abs/2019MNRAS.486.2075B}
}

@article{Gray1994_Spectrum,
  author   = {{Gray}, R.~O. and {Corbally}, C.~J.},
  title    = {{The Calibration of MK Spectral Classes Using Spectral Synthesis. I. The Effective Temperature Calibration of Dwarf Stars}},
  journal  = {\aj},
  year     = 1994,
  month    = feb,
  volume   = {107},
  pages    = {742},
  doi      = {10.1086/116893},
  adsurl   = {https://ui.adsabs.harvard.edu/abs/1994AJ....107..742G}
}

@inproceedings{Castelli2003_ATLAS9,
  author        = {{Castelli}, F. and {Kurucz}, R.~L.},
  title         = {{New Grids of ATLAS9 Model Atmospheres}},
  booktitle     = {Modelling of Stellar Atmospheres},
  year          = 2003,
  editor        = {{Piskunov}, N. and {Weiss}, W.~W. and {Gray}, D.~F.},
  series        = {IAU Symposium},
  volume        = {210},
  month         = jan,
  pages         = {A20},
  doi           = {10.48550/arXiv.astro-ph/0405087},
  archiveprefix = {arXiv},
  eprint        = {astro-ph/0405087},
  primaryclass  = {astro-ph},
  adsurl        = {https://ui.adsabs.harvard.edu/abs/2003IAUS..210P.A20C}
}

@article{Grevesse1998SS_solar_composition,
  author   = {{Grevesse}, N. and {Sauval}, A.~J.},
  title    = {{Standard Solar Composition}},
  journal  = {\ssr},
  year     = 1998,
  month    = may,
  volume   = {85},
  pages    = {161-174},
  doi      = {10.1023/A:1005161325181},
  adsurl   = {https://ui.adsabs.harvard.edu/abs/1998SSRv...85..161G}
}

@article{Ryabchikova2015_VALD3,
  author  = {{Ryabchikova}, T. and {Piskunov}, N. and {Kurucz}, R.~L. and {Stempels}, H.~C. and {Heiter}, U. and {Pakhomov}, Yu and {Barklem}, P.~S.},
  title   = {{A major upgrade of the VALD database}},
  journal = {\physscr},
  year    = 2015,
  month   = may,
  volume  = {90},
  number  = {5},
  eid     = {054005},
  pages   = {054005},
  doi     = {10.1088/0031-8949/90/5/054005},
  adsurl  = {https://ui.adsabs.harvard.edu/abs/2015PhyS...90e4005R}
}

@book{Gray2005_OASP,
  author  = {{Gray}, David F.},
  title   = {{The Observation and Analysis of Stellar Photospheres}},
  year    = 2005,
  doi     = {10.1017/CBO9781316036570},
  adsurl  = {https://ui.adsabs.harvard.edu/abs/2005oasp.book.....G}
}

@software{Morton2015_isochrones,
  author       = {{Morton}, Timothy D.},
  title        = {{isochrones: Stellar model grid package}},
  howpublished = {Astrophysics Source Code Library, record ascl:1503.010},
  year         = 2015,
  month        = mar,
  eid          = {ascl:1503.010},
  adsurl       = {https://ui.adsabs.harvard.edu/abs/2015ascl.soft03010M}
}

@article{Choi2016_MIST,
  author        = {{Choi}, Jieun and {Dotter}, Aaron and {Conroy}, Charlie and {Cantiello}, Matteo and {Paxton}, Bill and {Johnson}, Benjamin D.},
  title         = {{Mesa Isochrones and Stellar Tracks (MIST). I. Solar-scaled Models}},
  journal       = {\apj},
  year          = 2016,
  month         = jun,
  volume        = {823},
  number        = {2},
  eid           = {102},
  pages         = {102},
  doi           = {10.3847/0004-637X/823/2/102},
  archiveprefix = {arXiv},
  eprint        = {1604.08592},
  primaryclass  = {astro-ph.SR},
  adsurl        = {https://ui.adsabs.harvard.edu/abs/2016ApJ...823..102C}
}

@article{Dotter2016_MIST,
  author        = {{Dotter}, Aaron},
  title         = {{MESA Isochrones and Stellar Tracks (MIST) 0: Methods for the Construction of Stellar Isochrones}},
  journal       = {\apjs},
  year          = 2016,
  month         = jan,
  volume        = {222},
  number        = {1},
  eid           = {8},
  pages         = {8},
  doi           = {10.3847/0067-0049/222/1/8},
  archiveprefix = {arXiv},
  eprint        = {1601.05144},
  primaryclass  = {astro-ph.SR},
  adsurl        = {https://ui.adsabs.harvard.edu/abs/2016ApJS..222....8D}
}

@article{Cardelli1989_extinction,
  author   = {{Cardelli}, Jason A. and {Clayton}, Geoffrey C. and {Mathis}, John S.},
  title    = {{The Relationship between Infrared, Optical, and Ultraviolet Extinction}},
  journal  = {\apj},
  year     = 1989,
  month    = oct,
  volume   = {345},
  pages    = {245},
  doi      = {10.1086/167900},
  adsurl   = {https://ui.adsabs.harvard.edu/abs/1989ApJ...345..245C}
}

@article{Zhu2022_Two,
  title         = {Two {{Candidate KH 15D-like Systems}} from the {{Zwicky Transient Facility}}},
  author        = {Zhu, Wei and Bernhard, Klaus and Dai, Fei and Fang, Min and Zanazzi, J. J. and Zang, Weicheng and Dong, Subo and Hambsch, Franz-Josef and Gan, Tianjun and Wu, Zexuan and Poon, Michael},
  year          = {2022},
  month         = jul,
  journal       = {The Astrophysical Journal Letters},
  volume        = {933},
  number        = {1},
  eprint        = {2206.00813},
  primaryclass  = {astro-ph},
  pages         = {L21},
  issn          = {2041-8205, 2041-8213},
  doi           = {10.3847/2041-8213/ac7b2d},
  urldate       = {2022-09-02},
  archiveprefix = {arxiv}
}

@article{Foreman-Mackey2013_emcee,
  title      = {Emcee: {{The MCMC Hammer}}},
  shorttitle = {Emcee},
  author     = {{Foreman-Mackey}, Daniel and Hogg, David W. and Lang, Dustin and Goodman, Jonathan},
  year       = {2013},
  month      = mar,
  journal    = {Publications of the Astronomical Society of the Pacific},
  volume     = {125},
  pages      = {306},
  issn       = {0004-6280},
  doi        = {10.1086/670067}
}

@article{Xiang15_lamost_stellar,
  author        = {{Xiang}, M.~S. and {Liu}, X.~W. and {Yuan}, H.~B. and {Huang}, Y. and {Huo}, Z.~Y. and {Zhang}, H.~W. and {Chen}, B.~Q. and {Zhang}, H.~H. and {Sun}, N.~C. and {Wang}, C. and {Zhao}, Y.~H. and {Shi}, J.~R. and {Luo}, A.~L. and {Li}, G.~P. and {Wu}, Y. and {Bai}, Z.~R. and {Zhang}, Y. and {Hou}, Y.~H. and {Yuan}, H.~L. and {Li}, G.~W. and {Wei}, Z.},
  title         = {{The LAMOST stellar parameter pipeline at Peking University - LSP3}},
  journal       = {\mnras},
  year          = 2015,
  month         = mar,
  volume        = {448},
  number        = {1},
  pages         = {822-854},
  doi           = {10.1093/mnras/stu2692},
  archiveprefix = {arXiv},
  eprint        = {1412.6627},
  primaryclass  = {astro-ph.GA},
  adsurl        = {https://ui.adsabs.harvard.edu/abs/2015MNRAS.448..822X}
}

@article{Aguado21_s2_stream,
  author        = {{Aguado}, David S. and {Myeong}, G.~C. and {Belokurov}, Vasily and {Evans}, N. Wyn and {Koposov}, Sergey E. and {Allende Prieto}, Carlos and {Lanfranchi}, Gustavo A. and {Matteucci}, Francesca and {Shetrone}, Matthew and {Sbordone}, Luca and {Navarrete}, Camila and {Gonz{\'a}lez Hern{\'a}ndez}, Jonay I. and {Chanam{\'e}}, Julio and {Peralta de Arriba}, Luis and {Yuan}, Zhen},
  title         = {{The S2 stream: the shreds of a primitive dwarf galaxy}},
  journal       = {\mnras},
  year          = 2021,
  month         = jan,
  volume        = {500},
  number        = {1},
  pages         = {889-910},
  doi           = {10.1093/mnras/staa3250},
  archiveprefix = {arXiv},
  eprint        = {2007.11003},
  primaryclass  = {astro-ph.GA},
  adsurl        = {https://ui.adsabs.harvard.edu/abs/2021MNRAS.500..889A}
}

@article{Husser2013_phoenix,
  author        = {{Husser}, T.-O. and {Wende-von Berg}, S. and {Dreizler}, S. and {Homeier}, D. and {Reiners}, A. and {Barman}, T. and {Hauschildt}, P.~H.},
  title         = {{A new extensive library of PHOENIX stellar atmospheres and synthetic spectra}},
  journal       = {\aap},
  year          = 2013,
  month         = may,
  volume        = {553},
  eid           = {A6},
  pages         = {A6},
  doi           = {10.1051/0004-6361/201219058},
  archiveprefix = {arXiv},
  eprint        = {1303.5632},
  primaryclass  = {astro-ph.SR},
  adsurl        = {https://ui.adsabs.harvard.edu/abs/2013A&A...553A...6H}
}

@article{Fulton18_radvel_radial,
  author        = {{Fulton}, Benjamin J. and {Petigura}, Erik A. and {Blunt}, Sarah and {Sinukoff}, Evan},
  title         = {{RadVel: The Radial Velocity Modeling Toolkit}},
  journal       = {\pasp},
  year          = 2018,
  month         = apr,
  volume        = {130},
  number        = {986},
  pages         = {044504},
  doi           = {10.1088/1538-3873/aaaaa8},
  archiveprefix = {arXiv},
  eprint        = {1801.01947},
  primaryclass  = {astro-ph.IM},
  adsurl        = {https://ui.adsabs.harvard.edu/abs/2018PASP..130d4504F}
}

@article{Winn06_orbit_occultations,
  author        = {{Winn}, Joshua N. and {Hamilton}, Catrina M. and {Herbst}, William J. and {Hoffman}, Jennifer L. and {Holman}, Matthew J. and {Johnson}, John A. and {Kuchner}, Marc J.},
  title         = {{The Orbit and Occultations of KH 15D}},
  journal       = {\apj},
  year          = 2006,
  month         = jun,
  volume        = {644},
  number        = {1},
  pages         = {510-524},
  doi           = {10.1086/503417},
  archiveprefix = {arXiv},
  eprint        = {astro-ph/0602352},
  primaryclass  = {astro-ph},
  adsurl        = {https://ui.adsabs.harvard.edu/abs/2006ApJ...644..510W}
}

@article{Masci19_zwicky_transient,
  author        = {{Masci}, Frank J. and {Laher}, Russ R. and {Rusholme}, Ben and {Shupe}, David L. and {Groom}, Steven and {Surace}, Jason and {Jackson}, Edward and {Monkewitz}, Serge and {Beck}, Ron and {Flynn}, David and {Terek}, Scott and {Landry}, Walter and {Hacopians}, Eugean and {Desai}, Vandana and {Howell}, Justin and {Brooke}, Tim and {Imel}, David and {Wachter}, Stefanie and {Ye}, Quan-Zhi and {Lin}, Hsing-Wen and {Cenko}, S. Bradley and {Cunningham}, Virginia and {Rebbapragada}, Umaa and {Bue}, Brian and {Miller}, Adam A. and {Mahabal}, Ashish and {Bellm}, Eric C. and {Patterson}, Maria T. and {Juri{\'c}}, Mario and {Golkhou}, V. Zach and {Ofek}, Eran O. and {Walters}, Richard and {Graham}, Matthew and {Kasliwal}, Mansi M. and {Dekany}, Richard G. and {Kupfer}, Thomas and {Burdge}, Kevin and {Cannella}, Christopher B. and {Barlow}, Tom and {Van Sistine}, Angela and {Giomi}, Matteo and {Fremling}, Christoffer and {Blagorodnova}, Nadejda and {Levitan}, David and {Riddle}, Reed and {Smith}, Roger M. and {Helou}, George and {Prince}, Thomas A. and {Kulkarni}, Shrinivas R.},
  title         = {{The Zwicky Transient Facility: Data Processing, Products, and Archive}},
  journal       = {\pasp},
  year          = 2019,
  month         = jan,
  volume        = {131},
  number        = {995},
  pages         = {018003},
  doi           = {10.1088/1538-3873/aae8ac},
  archiveprefix = {arXiv},
  eprint        = {1902.01872},
  primaryclass  = {astro-ph.IM},
  adsurl        = {https://ui.adsabs.harvard.edu/abs/2019PASP..131a8003M}
}

@article{Hunt24_improving_open,
  author        = {{Hunt}, Emily L. and {Reffert}, Sabine},
  title         = {{Improving the open cluster census. III. Using cluster masses, radii, and dynamics to create a cleaned open cluster catalogue}},
  journal       = {\aap},
  year          = 2024,
  month         = jun,
  volume        = {686},
  eid           = {A42},
  pages         = {A42},
  doi           = {10.1051/0004-6361/202348662},
  archiveprefix = {arXiv},
  eprint        = {2403.05143},
  primaryclass  = {astro-ph.GA},
  adsurl        = {https://ui.adsabs.harvard.edu/abs/2024A&A...686A..42H}
}

@article{Jeffries23_gaia-eso_survey,
  author        = {{Jeffries}, R.~D. and {Jackson}, R.~J. and {Wright}, Nicholas J. and {Weaver}, G. and {Gilmore}, G. and {Randich}, S. and {Bragaglia}, A. and {Korn}, A.~J. and {Smiljanic}, R. and {Biazzo}, K. and {Casey}, A.~R. and {Frasca}, A. and {Gonneau}, A. and {Guiglion}, G. and {Morbidelli}, L. and {Prisinzano}, L. and {Sacco}, G.~G. and {Tautvai{\v{s}}ien{\.{e}}}, G. and {Worley}, C.~C. and {Zaggia}, S.},
  title         = {{The Gaia-ESO Survey: empirical estimates of stellar ages from lithium equivalent widths (EAGLES)}},
  journal       = {\mnras},
  year          = 2023,
  month         = jul,
  volume        = {523},
  number        = {1},
  pages         = {802-824},
  doi           = {10.1093/mnras/stad1293},
  archiveprefix = {arXiv},
  eprint        = {2304.12197},
  primaryclass  = {astro-ph.SR},
  adsurl        = {https://ui.adsabs.harvard.edu/abs/2023MNRAS.523..802J}
}

@article{Hu26_six,
  author        = {{Hu}, Zhecheng and {Zhu}, Wei and {Wang}, Shuming and {Wang}, Sharon Xuesong},
  title         = {{Six New Circumbinary Disk Occultation Candidates from the Zwicky Transient Facility}},
  journal       = {\apjs},
  year          = 2026,
  month         = jun,
  volume        = {284},
  number        = {2},
  eid           = {70},
  pages         = {70},
  doi           = {10.3847/1538-4365/ae5d41},
  archiveprefix = {arXiv},
  eprint        = {2601.16828},
  primaryclass  = {astro-ph.SR},
  adsurl        = {https://ui.adsabs.harvard.edu/abs/2026ApJS..284...70H}
}

@article{Rossiter1924,
  author  = {{Rossiter}, R.~A.},
  title   = {{On the detection of an effect of rotation during eclipse in the velocity of the brighter component of beta Lyrae, and on the constancy of velocity of this system.}},
  journal = {\apj},
  year    = 1924,
  month   = jul,
  volume  = {60},
  pages   = {15-21},
  doi     = {10.1086/142825},
  adsurl  = {https://ui.adsabs.harvard.edu/abs/1924ApJ....60...15R}
}

@article{McLaughlin1924,
  author  = {{McLaughlin}, D.~B.},
  title   = {{Some results of a spectrographic study of the Algol system.}},
  journal = {\apj},
  year    = 1924,
  month   = jul,
  volume  = {60},
  pages   = {22-31},
  doi     = {10.1086/142826},
  adsurl  = {https://ui.adsabs.harvard.edu/abs/1924ApJ....60...22M}
}

@article{Poon21_constraining_circumbinary,
  author        = {{Poon}, Michael and {Zanazzi}, J.~J. and {Zhu}, Wei},
  title         = {{Constraining the circumbinary disc tilt in the KH 15D system}},
  journal       = {\mnras},
  year          = 2021,
  month         = may,
  volume        = {503},
  number        = {2},
  pages         = {1599-1614},
  doi           = {10.1093/mnras/stab575},
  archiveprefix = {arXiv},
  eprint        = {2009.14204},
  primaryclass  = {astro-ph.EP},
  adsurl        = {https://ui.adsabs.harvard.edu/abs/2021MNRAS.503.1599P}
}

@article{GarciaSoto2020,
  author   = {{Garc{\'\i}a Soto}, Aylin and {Ali}, Aleezah and {Newmark}, Amanda and {Herbst}, William and {Windemuth}, Diana and {Winn}, Joshua N.},
  title    = {{Evidence for Transparency and Clumps in the Circumbinary Ring of the T Tauri Star V582 Mon (KH 15D)}},
  journal  = {\aj},
  year     = 2020,
  month    = apr,
  volume   = {159},
  number   = {4},
  eid      = {135},
  pages    = {135},
  doi      = {10.3847/1538-3881/ab6efd},
  adsurl   = {https://ui.adsabs.harvard.edu/abs/2020AJ....159..135G}
}

@article{Arulanantham16_seeing_through,
  author        = {{Arulanantham}, Nicole A. and {Herbst}, William and {Cody}, Ann Marie and {Stauffer}, John R. and {Rebull}, Luisa M. and {Agol}, Eric and {Windemuth}, Diana and {Marengo}, Massimo and {Winn}, Joshua N. and {Hamilton}, Catrina M. and {Mundt}, Reinhard and {Johns-Krull}, Christopher M. and {Gutermuth}, Robert A.},
  title         = {{Seeing Through the Ring: Near-infrared Photometry of V582 Mon (KH 15D)}},
  journal       = {\aj},
  year          = 2016,
  month         = apr,
  volume        = {151},
  number        = {4},
  eid           = {90},
  pages         = {90},
  doi           = {10.3847/0004-6256/151/4/90},
  archiveprefix = {arXiv},
  eprint        = {1602.01877},
  primaryclass  = {astro-ph.SR},
  adsurl        = {https://ui.adsabs.harvard.edu/abs/2016AJ....151...90A}
}

@article{Tofflemire17_accretion_magnetic,
  author        = {{Tofflemire}, Benjamin M. and {Mathieu}, Robert D. and {Ardila}, David R. and {Akeson}, Rachel L. and {Ciardi}, David R. and {Johns-Krull}, Christopher and {Herczeg}, Gregory J. and {Quijano-Vodniza}, Alberto},
  title         = {{Accretion and Magnetic Reconnection in the Classical T Tauri Binary DQ Tau}},
  journal       = {\apj},
  year          = 2017,
  month         = jan,
  volume        = {835},
  number        = {1},
  eid           = {8},
  pages         = {8},
  doi           = {10.3847/1538-4357/835/1/8},
  archiveprefix = {arXiv},
  eprint        = {1612.02431},
  primaryclass  = {astro-ph.SR},
  adsurl        = {https://ui.adsabs.harvard.edu/abs/2017ApJ...835....8T}
}

@article{Hamilton12_complex_variability,
  author        = {{Hamilton}, Catrina M. and {Johns-Krull}, Christopher M. and {Mundt}, Reinhard and {Herbst}, William and {Winn}, Joshua N.},
  title         = {{Complex Variability of the H{\ensuremath{\alpha}} Emission Line Profile of the T Tauri Binary System KH 15D: The Influence of Orbital Phase, Occultation by the Circumbinary Disk, and Accretion Phenomena}},
  journal       = {\apj},
  year          = 2012,
  month         = jun,
  volume        = {751},
  number        = {2},
  eid           = {147},
  pages         = {147},
  doi           = {10.1088/0004-637X/751/2/147},
  archiveprefix = {arXiv},
  eprint        = {1204.1334},
  primaryclass  = {astro-ph.SR},
  adsurl        = {https://ui.adsabs.harvard.edu/abs/2012ApJ...751..147H}
}

@article{Alcala14_x-shooter_spectroscopy,
  author        = {{Alcal{\'a}}, J.~M. and {Natta}, A. and {Manara}, C.~F. and {Spezzi}, L. and {Stelzer}, B. and {Frasca}, A. and {Biazzo}, K. and {Covino}, E. and {Randich}, S. and {Rigliaco}, E. and {Testi}, L. and {Comer{\'o}n}, F. and {Cupani}, G. and {D'Elia}, V.},
  title         = {{X-shooter spectroscopy of young stellar objects. IV. Accretion in low-mass stars and substellar objects in Lupus}},
  journal       = {\aap},
  year          = 2014,
  month         = jan,
  volume        = {561},
  eid           = {A2},
  pages         = {A2},
  doi           = {10.1051/0004-6361/201322254},
  archiveprefix = {arXiv},
  eprint        = {1310.2069},
  primaryclass  = {astro-ph.SR},
  adsurl        = {https://ui.adsabs.harvard.edu/abs/2014A&A...561A...2A}
}

@article{Erkal22_he_i,
  author        = {{Erkal}, J. and {Manara}, C.~F. and {Schneider}, P.~C. and {Vincenzi}, M. and {Nisini}, B. and {Coffey}, D. and {Alcal{\'a}}, J.~M. and {Fedele}, D. and {Antoniucci}, S.},
  title         = {{The He I {\ensuremath{\lambda}}10830 {\r{A}} line as a probe of winds and accretion in young stars in Lupus and Upper Scorpius}},
  journal       = {\aap},
  year          = 2022,
  month         = oct,
  volume        = {666},
  eid           = {A188},
  pages         = {A188},
  doi           = {10.1051/0004-6361/202244254},
  archiveprefix = {arXiv},
  eprint        = {2208.02940},
  primaryclass  = {astro-ph.SR},
  adsurl        = {https://ui.adsabs.harvard.edu/abs/2022A&A...666A.188E}
}

@article{Edwards06_probing_t,
  author        = {{Edwards}, Suzan and {Fischer}, William and {Hillenbrand}, Lynne and {Kwan}, John},
  title         = {{Probing T Tauri Accretion and Outflow with 1 Micron Spectroscopy}},
  journal       = {\apj},
  year          = 2006,
  month         = jul,
  volume        = {646},
  number        = {1},
  pages         = {319-341},
  doi           = {10.1086/504832},
  archiveprefix = {arXiv},
  eprint        = {astro-ph/0604006},
  primaryclass  = {astro-ph},
  adsurl        = {https://ui.adsabs.harvard.edu/abs/2006ApJ...646..319E}
}

@article{Kwan2007_modeling_t,
  author        = {{Kwan}, John and {Edwards}, Suzan and {Fischer}, William},
  title         = {{Modeling T Tauri Winds from He I {\ensuremath{\lambda}}10830 Profiles}},
  journal       = {\apj},
  year          = 2007,
  month         = mar,
  volume        = {657},
  number        = {2},
  pages         = {897-915},
  doi           = {10.1086/511057},
  archiveprefix = {arXiv},
  eprint        = {astro-ph/0611585},
  primaryclass  = {astro-ph},
  adsurl        = {https://ui.adsabs.harvard.edu/abs/2007ApJ...657..897K}
}

@article{Miranda2015_tidal_truncation,
  author        = {{Miranda}, Ryan and {Lai}, Dong},
  title         = {{Tidal truncation of inclined circumstellar and circumbinary discs in young stellar binaries}},
  journal       = {\mnras},
  year          = 2015,
  month         = sep,
  volume        = {452},
  number        = {3},
  pages         = {2396-2409},
  doi           = {10.1093/mnras/stv1450},
  archiveprefix = {arXiv},
  eprint        = {1504.02917},
  primaryclass  = {astro-ph.EP},
  adsurl        = {https://ui.adsabs.harvard.edu/abs/2015MNRAS.452.2396M}
}

@article{Smallwood23_formation_polar,
  author        = {{Smallwood}, Jeremy L. and {Martin}, Rebecca G. and {Lubow}, Stephen H.},
  title         = {{Formation of polar circumstellar discs in binary star systems}},
  journal       = {\mnras},
  year          = 2023,
  month         = apr,
  volume        = {520},
  number        = {2},
  pages         = {2952-2964},
  doi           = {10.1093/mnras/stad338},
  archiveprefix = {arXiv},
  eprint        = {2301.11769},
  primaryclass  = {astro-ph.EP},
  adsurl        = {https://ui.adsabs.harvard.edu/abs/2023MNRAS.520.2952S}
}

@article{Artymowicz96_mass_flow,
  author   = {{Artymowicz}, Pawel and {Lubow}, Stephen H.},
  title    = {{Mass Flow through Gaps in Circumbinary Disks}},
  journal  = {\apjl},
  year     = 1996,
  month    = aug,
  volume   = {467},
  pages    = {L77},
  doi      = {10.1086/310200},
  adsurl   = {https://ui.adsabs.harvard.edu/abs/1996ApJ...467L..77A}
}

@article{Basri97_classical_t,
  author   = {{Basri}, Gibor and {Johns-Krull}, Christopher M. and {Mathieu}, Robert D.},
  title    = {{The Classical T Tauri Spectroscopic Binary DQ Tau. II. Emission Line Variations with Orbital Phase.}},
  journal  = {\aj},
  year     = 1997,
  month    = aug,
  volume   = {114},
  pages    = {781-792},
  doi      = {10.1086/118510},
  adsurl   = {https://ui.adsabs.harvard.edu/abs/1997AJ....114..781B}
}

@article{Munoz16_pulsed_accretion,
  author        = {{Mu{\~n}oz}, Diego J. and {Lai}, Dong},
  title         = {{Pulsed Accretion onto Eccentric and Circular Binaries}},
  journal       = {\apj},
  year          = 2016,
  month         = aug,
  volume        = {827},
  number        = {1},
  eid           = {43},
  pages         = {43},
  doi           = {10.3847/0004-637X/827/1/43},
  archiveprefix = {arXiv},
  eprint        = {1604.00004},
  primaryclass  = {astro-ph.EP},
  adsurl        = {https://ui.adsabs.harvard.edu/abs/2016ApJ...827...43M}
}

@article{Smallwood22_accretion_binary,
  author        = {{Smallwood}, Jeremy L. and {Lubow}, Stephen H. and {Martin}, Rebecca G.},
  title         = {{Accretion on to a binary from a polar circumbinary disc}},
  journal       = {\mnras},
  year          = 2022,
  month         = jul,
  volume        = {514},
  number        = {1},
  pages         = {1249-1257},
  doi           = {10.1093/mnras/stac1416},
  archiveprefix = {arXiv},
  eprint        = {2205.09183},
  primaryclass  = {astro-ph.EP},
  adsurl        = {https://ui.adsabs.harvard.edu/abs/2022MNRAS.514.1249S}
}

@article{Ragusa20_evolution_large,
  author        = {{Ragusa}, Enrico and {Alexander}, Richard and {Calcino}, Josh and {Hirsh}, Kieran and {Price}, Daniel J.},
  title         = {{The evolution of large cavities and disc eccentricity in circumbinary discs}},
  journal       = {\mnras},
  year          = 2020,
  month         = dec,
  volume        = {499},
  number        = {3},
  pages         = {3362-3380},
  doi           = {10.1093/mnras/staa2954},
  archiveprefix = {arXiv},
  eprint        = {2009.10738},
  primaryclass  = {astro-ph.EP},
  adsurl        = {https://ui.adsabs.harvard.edu/abs/2020MNRAS.499.3362R}
}

@article{Raghavan10_survey_stellar,
  author        = {{Raghavan}, Deepak and {McAlister}, Harold A. and {Henry}, Todd J. and {Latham}, David W. and {Marcy}, Geoffrey W. and {Mason}, Brian D. and {Gies}, Douglas R. and {White}, Russel J. and {ten Brummelaar}, Theo A.},
  title         = {{A Survey of Stellar Families: Multiplicity of Solar-type Stars}},
  journal       = {\apjs},
  year          = 2010,
  month         = sep,
  volume        = {190},
  number        = {1},
  pages         = {1-42},
  doi           = {10.1088/0067-0049/190/1/1},
  archiveprefix = {arXiv},
  eprint        = {1007.0414},
  primaryclass  = {astro-ph.SR},
  adsurl        = {https://ui.adsabs.harvard.edu/abs/2010ApJS..190....1R}
}

@article{Duchene13_stellar_multiplicity,
  author        = {{Duch{\^e}ne}, Gaspard and {Kraus}, Adam},
  title         = {{Stellar Multiplicity}},
  journal       = {\araa},
  year          = 2013,
  month         = aug,
  volume        = {51},
  number        = {1},
  pages         = {269-310},
  doi           = {10.1146/annurev-astro-081710-102602},
  archiveprefix = {arXiv},
  eprint        = {1303.3028},
  primaryclass  = {astro-ph.SR},
  adsurl        = {https://ui.adsabs.harvard.edu/abs/2013ARA&A..51..269D}
}

@article{Kearns98_additional_periodic,
  author   = {{Kearns}, Kristin E. and {Herbst}, William},
  title    = {{Additional Periodic Variables in NGC 2264}},
  journal  = {\aj},
  year     = 1998,
  month    = jul,
  volume   = {116},
  number   = {1},
  pages    = {261-265},
  doi      = {10.1086/300426},
  adsurl   = {https://ui.adsabs.harvard.edu/abs/1998AJ....116..261K}
}

@article{Chiang04_circumbinary_ring,
  author        = {{Chiang}, Eugene I. and {Murray-Clay}, Ruth A.},
  title         = {{The Circumbinary Ring of KH 15D}},
  journal       = {\apj},
  year          = 2004,
  month         = jun,
  volume        = {607},
  number        = {2},
  pages         = {913-920},
  doi           = {10.1086/383522},
  archiveprefix = {arXiv},
  eprint        = {astro-ph/0312515},
  primaryclass  = {astro-ph},
  adsurl        = {https://ui.adsabs.harvard.edu/abs/2004ApJ...607..913C}
}

@article{Winn04_kh_15d,
  author        = {{Winn}, Joshua N. and {Holman}, Matthew J. and {Johnson}, John A. and {Stanek}, Krzysztof Z. and {Garnavich}, Peter M.},
  title         = {{KH 15D: Gradual Occultation of a Pre-Main-Sequence Binary}},
  journal       = {\apjl},
  year          = 2004,
  month         = mar,
  volume        = {603},
  number        = {1},
  pages         = {L45-L48},
  doi           = {10.1086/383089},
  archiveprefix = {arXiv},
  eprint        = {astro-ph/0312458},
  primaryclass  = {astro-ph},
  adsurl        = {https://ui.adsabs.harvard.edu/abs/2004ApJ...603L..45W}
}

@article{Bellm19_zwicky_transient,
  author        = {{Bellm}, Eric C. and {Kulkarni}, Shrinivas R. and {Graham}, Matthew J. and {Dekany}, Richard and {Smith}, Roger M. and {Riddle}, Reed and {Masci}, Frank J. and {Helou}, George and {Prince}, Thomas A. and {Adams}, Scott M. and {Barbarino}, C. and {Barlow}, Tom and {Bauer}, James and {Beck}, Ron and {Belicki}, Justin and {Biswas}, Rahul and {Blagorodnova}, Nadejda and {Bodewits}, Dennis and {Bolin}, Bryce and {Brinnel}, Valery and {Brooke}, Tim and {Bue}, Brian and {Bulla}, Mattia and {Burruss}, Rick and {Cenko}, S. Bradley and {Chang}, Chan-Kao and {Connolly}, Andrew and {Coughlin}, Michael and {Cromer}, John and {Cunningham}, Virginia and {De}, Kishalay and {Delacroix}, Alex and {Desai}, Vandana and {Duev}, Dmitry A. and {Eadie}, Gwendolyn and {Farnham}, Tony L. and {Feeney}, Michael and {Feindt}, Ulrich and {Flynn}, David and {Franckowiak}, Anna and {Frederick}, S. and {Fremling}, C. and {Gal-Yam}, Avishay and {Gezari}, Suvi and {Giomi}, Matteo and {Goldstein}, Daniel A. and {Golkhou}, V. Zach and {Goobar}, Ariel and {Groom}, Steven and {Hacopians}, Eugean and {Hale}, David and {Henning}, John and {Ho}, Anna Y.~Q. and {Hover}, David and {Howell}, Justin and {Hung}, Tiara and {Huppenkothen}, Daniela and {Imel}, David and {Ip}, Wing-Huen and {Ivezi{\'c}}, {\v{Z}}eljko and {Jackson}, Edward and {Jones}, Lynne and {Juric}, Mario and {Kasliwal}, Mansi M. and {Kaspi}, S. and {Kaye}, Stephen and {Kelley}, Michael S.~P. and {Kowalski}, Marek and {Kramer}, Emily and {Kupfer}, Thomas and {Landry}, Walter and {Laher}, Russ R. and {Lee}, Chien-De and {Lin}, Hsing Wen and {Lin}, Zhong-Yi and {Lunnan}, Ragnhild and {Giomi}, Matteo and {Mahabal}, Ashish and {Mao}, Peter and {Miller}, Adam A. and {Monkewitz}, Serge and {Murphy}, Patrick and {Ngeow}, Chow-Choong and {Nordin}, Jakob and {Nugent}, Peter and {Ofek}, Eran and {Patterson}, Maria T. and {Penprase}, Bryan and {Porter}, Michael and {Rauch}, Ludwig and {Rebbapragada}, Umaa and {Reiley}, Dan and {Rigault}, Mickael and {Rodriguez}, Hector and {van Roestel}, Jan and {Rusholme}, Ben and {van Santen}, Jakob and {Schulze}, S. and {Shupe}, David L. and {Singer}, Leo P. and {Soumagnac}, Maayane T. and {Stein}, Robert and {Surace}, Jason and {Sollerman}, Jesper and {Szkody}, Paula and {Taddia}, F. and {Terek}, Scott and {Van Sistine}, Angela and {van Velzen}, Sjoert and {Vestrand}, W. Thomas and {Walters}, Richard and {Ward}, Charlotte and {Ye}, Quan-Zhi and {Yu}, Po-Chieh and {Yan}, Lin and {Zolkower}, Jeffry},
  title         = {{The Zwicky Transient Facility: System Overview, Performance, and First Results}},
  journal       = {\pasp},
  year          = 2019,
  month         = jan,
  volume        = {131},
  number        = {995},
  pages         = {018002},
  doi           = {10.1088/1538-3873/aaecbe},
  archiveprefix = {arXiv},
  eprint        = {1902.01932},
  primaryclass  = {astro-ph.IM},
  adsurl        = {https://ui.adsabs.harvard.edu/abs/2019PASP..131a8002B}
}

@article{Udalski97_optical_gravitational,
  author        = {{Udalski}, A. and {Kubiak}, M. and {Szymanski}, M.},
  title         = {{Optical Gravitational Lensing Experiment. OGLE-2 -- the Second Phase of the OGLE Project}},
  journal       = {\actaa},
  year          = 1997,
  month         = jul,
  volume        = {47},
  pages         = {319-344},
  doi           = {10.48550/arXiv.astro-ph/9710091},
  archiveprefix = {arXiv},
  eprint        = {astro-ph/9710091},
  primaryclass  = {astro-ph},
  adsurl        = {https://ui.adsabs.harvard.edu/abs/1997AcA....47..319U}
}

@article{Udalski03_optical_gravitational,
  author        = {{Udalski}, A.},
  title         = {{The Optical Gravitational Lensing Experiment. Real Time Data Analysis Systems in the OGLE-III Survey}},
  journal       = {\actaa},
  year          = 2003,
  month         = dec,
  volume        = {53},
  pages         = {291-305},
  doi           = {10.48550/arXiv.astro-ph/0401123},
  archiveprefix = {arXiv},
  eprint        = {astro-ph/0401123},
  primaryclass  = {astro-ph},
  adsurl        = {https://ui.adsabs.harvard.edu/abs/2003AcA....53..291U}
}

@article{Udalski15_ogle-iv_fourth,
  author        = {{Udalski}, A. and {Szyma{\'n}ski}, M.~K. and {Szyma{\'n}ski}, G.},
  title         = {{OGLE-IV: Fourth Phase of the Optical Gravitational Lensing Experiment}},
  journal       = {\actaa},
  year          = 2015,
  month         = mar,
  volume        = {65},
  number        = {1},
  pages         = {1-38},
  doi           = {10.48550/arXiv.1504.05966},
  archiveprefix = {arXiv},
  eprint        = {1504.05966},
  primaryclass  = {astro-ph.SR},
  adsurl        = {https://ui.adsabs.harvard.edu/abs/2015AcA....65....1U}
}

@article{Urbanowicz26_thirty_circumbinary,
  author        = {{Urbanowicz}, M.~A. and {Soszy{\'n}ski}, I. and {Pietrukowicz}, P. and {Udalski}, A. and {Mr{\'o}z}, P. and {Wrona}, M. and {Ratajczak}, M. and {Szyma{\'n}ski}, M.~K. and {Skowron}, J. and {Skowron}, D.~M. and {Poleski}, R. and {Koz{\l}owski}, S. and {Iwanek}, P. and {Ulaczyk}, K. and {Rybicki}, K. and {Gromadzki}, M. and {Mr{\'o}z}, M.},
  title         = {{Thirty Circumbinary Disk Occultation Systems (KH 15D-like stars) from the OGLE Project}},
  journal       = {arXiv e-prints},
  year          = 2026,
  month         = may,
  eid           = {arXiv:2605.08470},
  pages         = {arXiv:2605.08470},
  doi           = {10.48550/arXiv.2605.08470},
  archiveprefix = {arXiv},
  eprint        = {2605.08470},
  primaryclass  = {astro-ph.SR},
  adsurl        = {https://ui.adsabs.harvard.edu/abs/2026arXiv260508470U}
}

@article{Minniti10_vista_variables,
  author        = {{Minniti}, D. and {Lucas}, P.~W. and {Emerson}, J.~P. and {Saito}, R.~K. and {Hempel}, M. and {Pietrukowicz}, P. and {Ahumada}, A.~V. and {Alonso}, M.~V. and {Alonso-Garcia}, J. and {Arias}, J.~I. and {Bandyopadhyay}, R.~M. and {Barb{\'a}}, R.~H. and {Barbuy}, B. and {Bedin}, L.~R. and {Bica}, E. and {Borissova}, J. and {Bronfman}, L. and {Carraro}, G. and {Catelan}, M. and {Clari{\'a}}, J.~J. and {Cross}, N. and {de Grijs}, R. and {D{\'e}k{\'a}ny}, I. and {Drew}, J.~E. and {Fari{\~n}a}, C. and {Feinstein}, C. and {Fern{\'a}ndez Laj{\'u}s}, E. and {Gamen}, R.~C. and {Geisler}, D. and {Gieren}, W. and {Goldman}, B. and {Gonzalez}, O.~A. and {Gunthardt}, G. and {Gurovich}, S. and {Hambly}, N.~C. and {Irwin}, M.~J. and {Ivanov}, V.~D. and {Jord{\'a}n}, A. and {Kerins}, E. and {Kinemuchi}, K. and {Kurtev}, R. and {L{\'o}pez-Corredoira}, M. and {Maccarone}, T. and {Masetti}, N. and {Merlo}, D. and {Messineo}, M. and {Mirabel}, I.~F. and {Monaco}, L. and {Morelli}, L. and {Padilla}, N. and {Palma}, T. and {Parisi}, M.~C. and {Pignata}, G. and {Rejkuba}, M. and {Roman-Lopes}, A. and {Sale}, S.~E. and {Schreiber}, M.~R. and {Schr{\"o}der}, A.~C. and {Smith}, M. and {Sodr{\'e}}, Jr., L. and {Soto}, M. and {Tamura}, M. and {Tappert}, C. and {Thompson}, M.~A. and {Toledo}, I. and {Zoccali}, M. and {Pietrzynski}, G.},
  title         = {{VISTA Variables in the Via Lactea (VVV): The public ESO near-IR variability survey of the Milky Way}},
  journal       = {\na},
  year          = 2010,
  month         = jul,
  volume        = {15},
  number        = {5},
  pages         = {433-443},
  doi           = {10.1016/j.newast.2009.12.002},
  archiveprefix = {arXiv},
  eprint        = {0912.1056},
  primaryclass  = {astro-ph.GA},
  adsurl        = {https://ui.adsabs.harvard.edu/abs/2010NewA...15..433M}
}

@article{Lucas24_most_variable,
  author        = {{Lucas}, P.~W. and {Smith}, L.~C. and {Guo}, Z. and {Contreras Pe{\~n}a}, C. and {Minniti}, D. and {Miller}, N. and {Alonso-Garc{\'\i}a}, J. and {Catelan}, M. and {Borissova}, J. and {Saito}, R.~K. and {Kurtev}, R. and {Navarro}, M.~G. and {Morris}, C. and {Muthu}, H. and {Froebrich}, D. and {Ivanov}, V.~D. and {Bayo}, A. and {Caratti o Garatti}, A. and {Sanders}, J.~L.},
  title         = {{The most variable VVV sources: eruptive protostars, dipping giants in the nuclear disc and others}},
  journal       = {\mnras},
  year          = 2024,
  month         = feb,
  volume        = {528},
  number        = {2},
  pages         = {1789-1822},
  doi           = {10.1093/mnras/stad3929},
  archiveprefix = {arXiv},
  eprint        = {2401.14471},
  primaryclass  = {astro-ph.SR},
  adsurl        = {https://ui.adsabs.harvard.edu/abs/2024MNRAS.528.1789L}
}

@article{Martin19_bebop_radial-velocity,
  author        = {{Martin}, David V. and {Triaud}, Amaury H.~M.~J. and {Udry}, St{\'e}phane and {Marmier}, Maxime and {Maxted}, Pierre F.~L. and {Collier Cameron}, Andrew and {Hellier}, Coel and {Pepe}, Francesco and {Pollacco}, Don and {S{\'e}gransan}, Damien and {West}, Richard},
  title         = {{The BEBOP radial-velocity survey for circumbinary planets. I. Eight years of CORALIE observations of 47 single-line eclipsing binaries and abundance constraints on the masses of circumbinary planets}},
  journal       = {\aap},
  year          = 2019,
  month         = apr,
  volume        = {624},
  eid           = {A68},
  pages         = {A68},
  doi           = {10.1051/0004-6361/201833669},
  archiveprefix = {arXiv},
  eprint        = {1901.01627},
  primaryclass  = {astro-ph.EP},
  adsurl        = {https://ui.adsabs.harvard.edu/abs/2019A&A...624A..68M}
}

@article{Doyle11_kepler-16_transiting,
  author        = {{Doyle}, Laurance R. and {Carter}, Joshua A. and {Fabrycky}, Daniel C. and {Slawson}, Robert W. and {Howell}, Steve B. and {Winn}, Joshua N. and {Orosz}, Jerome A. and {P{\v{r}}sa}, Andrej and {Welsh}, William F. and {Quinn}, Samuel N. and {Latham}, David and {Torres}, Guillermo and {Buchhave}, Lars A. and {Marcy}, Geoffrey W. and {Fortney}, Jonathan J. and {Shporer}, Avi and {Ford}, Eric B. and {Lissauer}, Jack J. and {Ragozzine}, Darin and {Rucker}, Michael and {Batalha}, Natalie and {Jenkins}, Jon M. and {Borucki}, William J. and {Koch}, David and {Middour}, Christopher K. and {Hall}, Jennifer R. and {McCauliff}, Sean and {Fanelli}, Michael N. and {Quintana}, Elisa V. and {Holman}, Matthew J. and {Caldwell}, Douglas A. and {Still}, Martin and {Stefanik}, Robert P. and {Brown}, Warren R. and {Esquerdo}, Gilbert A. and {Tang}, Sumin and {Furesz}, Gabor and {Geary}, John C. and {Berlind}, Perry and {Calkins}, Michael L. and {Short}, Donald R. and {Steffen}, Jason H. and {Sasselov}, Dimitar and {Dunham}, Edward W. and {Cochran}, William D. and {Boss}, Alan and {Haas}, Michael R. and {Buzasi}, Derek and {Fischer}, Debra},
  title         = {{Kepler-16: A Transiting Circumbinary Planet}},
  journal       = {Science},
  year          = 2011,
  month         = sep,
  volume        = {333},
  number        = {6049},
  pages         = {1602},
  doi           = {10.1126/science.1210923},
  archiveprefix = {arXiv},
  eprint        = {1109.3432},
  primaryclass  = {astro-ph.EP},
  adsurl        = {https://ui.adsabs.harvard.edu/abs/2011Sci...333.1602D}
}

@article{Kostov20_toi-1338_tess,
  author        = {{Kostov}, Veselin B. and {Orosz}, Jerome A. and {Feinstein}, Adina D. and {Welsh}, William F. and {Cukier}, Wolf and {Haghighipour}, Nader and {Quarles}, Billy and {Martin}, David V. and {Montet}, Benjamin T. and {Torres}, Guillermo and {Triaud}, Amaury H.~M.~J. and {Barclay}, Thomas and {Boyd}, Patricia and {Briceno}, Cesar and {Cameron}, Andrew Collier and {Correia}, Alexandre C.~M. and {Gilbert}, Emily A. and {Gill}, Samuel and {Gillon}, Micha{\"e}l and {Haqq-Misra}, Jacob and {Hellier}, Coel and {Dressing}, Courtney and {Fabrycky}, Daniel C. and {Furesz}, Gabor and {Jenkins}, Jon M. and {Kane}, Stephen R. and {Kopparapu}, Ravi and {Hod{\v{z}}i{\'c}}, Vedad Kunovac and {Latham}, David W. and {Law}, Nicholas and {Levine}, Alan M. and {Li}, Gongjie and {Lintott}, Chris and {Lissauer}, Jack J. and {Mann}, Andrew W. and {Mazeh}, Tsevi and {Mardling}, Rosemary and {Maxted}, Pierre F.~L. and {Eisner}, Nora and {Pepe}, Francesco and {Pepper}, Joshua and {Pollacco}, Don and {Quinn}, Samuel N. and {Quintana}, Elisa V. and {Rowe}, Jason F. and {Ricker}, George and {Rose}, Mark E. and {Seager}, S. and {Santerne}, Alexandre and {S{\'e}gransan}, Damien and {Short}, Donald R. and {Smith}, Jeffrey C. and {Standing}, Matthew R. and {Tokovinin}, Andrei and {Trifonov}, Trifon and {Turner}, Oliver and {Twicken}, Joseph D. and {Udry}, St{\'e}phane and {Vanderspek}, Roland and {Winn}, Joshua N. and {Wolf}, Eric T. and {Ziegler}, Carl and {Ansorge}, Peter and {Barnet}, Frank and {Bergeron}, Joel and {Huten}, Marc and {Pappa}, Giuseppe and {van der Straeten}, Timo},
  title         = {{TOI-1338: TESS' First Transiting Circumbinary Planet}},
  journal       = {\aj},
  year          = 2020,
  month         = jun,
  volume        = {159},
  number        = {6},
  eid           = {253},
  pages         = {253},
  doi           = {10.3847/1538-3881/ab8a48},
  archiveprefix = {arXiv},
  eprint        = {2004.07783},
  primaryclass  = {astro-ph.EP},
  adsurl        = {https://ui.adsabs.harvard.edu/abs/2020AJ....159..253K}
}

@article{Goldberg23_5m_sub,
  author        = {{Goldberg}, Max and {Fabrycky}, Daniel and {Martin}, David V. and {Albrecht}, Simon and {Deeg}, Hans J. and {Nowak}, Grzegorz},
  title         = {{A 5M$_{Jup}$ non-transiting coplanar circumbinary planet around Kepler-1660AB}},
  journal       = {\mnras},
  year          = 2023,
  month         = nov,
  volume        = {525},
  number        = {3},
  pages         = {4628-4641},
  doi           = {10.1093/mnras/stad2568},
  archiveprefix = {arXiv},
  eprint        = {2308.09255},
  primaryclass  = {astro-ph.EP},
  adsurl        = {https://ui.adsabs.harvard.edu/abs/2023MNRAS.525.4628G}
}

@article{Martin14_planets_transiting,
  author        = {{Martin}, David V. and {Triaud}, Amaury H.~M.~J.},
  title         = {{Planets transiting non-eclipsing binaries}},
  journal       = {\aap},
  year          = 2014,
  month         = oct,
  volume        = {570},
  eid           = {A91},
  pages         = {A91},
  doi           = {10.1051/0004-6361/201323112},
  archiveprefix = {arXiv},
  eprint        = {1404.5360},
  primaryclass  = {astro-ph.EP},
  adsurl        = {https://ui.adsabs.harvard.edu/abs/2014A&A...570A..91M}
}

@article{Chen22_number_transits,
  author        = {{Chen}, Zirui and {Kipping}, David},
  title         = {{The number of transits per epoch for transiting misaligned circumbinary planets}},
  journal       = {\mnras},
  year          = 2022,
  month         = jul,
  volume        = {513},
  number        = {4},
  pages         = {5162-5173},
  doi           = {10.1093/mnras/stac1246},
  archiveprefix = {arXiv},
  eprint        = {2112.00966},
  primaryclass  = {astro-ph.EP},
  adsurl        = {https://ui.adsabs.harvard.edu/abs/2022MNRAS.513.5162C}
}

@article{Martin17_polar_alignment,
  author        = {{Martin}, Rebecca G. and {Lubow}, Stephen H.},
  title         = {{Polar Alignment of a Protoplanetary Disk around an Eccentric Binary}},
  journal       = {\apjl},
  year          = 2017,
  month         = feb,
  volume        = {835},
  number        = {2},
  eid           = {L28},
  pages         = {L28},
  doi           = {10.3847/2041-8213/835/2/L28},
  archiveprefix = {arXiv},
  eprint        = {1702.00545},
  primaryclass  = {astro-ph.EP},
  adsurl        = {https://ui.adsabs.harvard.edu/abs/2017ApJ...835L..28M}
}

@article{Zanazzi18_inclination_evolution,
  author   = {{Zanazzi}, J.~J. and {Lai}, Dong},
  title    = {{Inclination evolution of protoplanetary discs around eccentric binaries}},
  journal  = {\mnras},
  year     = 2018,
  month    = jan,
  volume   = {473},
  number   = {1},
  pages    = {603-615},
  doi      = {10.1093/mnras/stx2375},
  adsurl   = {https://ui.adsabs.harvard.edu/abs/2018MNRAS.473..603Z}
}

@article{Childs21_formation_polar,
  author        = {{Childs}, Anna C. and {Martin}, Rebecca G.},
  title         = {{Formation of Polar Terrestrial Circumbinary Planets}},
  journal       = {\apjl},
  year          = 2021,
  month         = oct,
  volume        = {920},
  number        = {1},
  eid           = {L8},
  pages         = {L8},
  doi           = {10.3847/2041-8213/ac2957},
  archiveprefix = {arXiv},
  eprint        = {2109.11653},
  primaryclass  = {astro-ph.EP},
  adsurl        = {https://ui.adsabs.harvard.edu/abs/2021ApJ...920L...8C}
}

@article{Kohler11_orbit_gg,
  author        = {{K{\"o}hler}, R.},
  title         = {{The orbit of GG Tauri A}},
  journal       = {\aap},
  year          = 2011,
  month         = jun,
  volume        = {530},
  eid           = {A126},
  pages         = {A126},
  doi           = {10.1051/0004-6361/201016327},
  archiveprefix = {arXiv},
  eprint        = {1104.2245},
  primaryclass  = {astro-ph.SR},
  adsurl        = {https://ui.adsabs.harvard.edu/abs/2011A&A...530A.126K}
}

@article{Kennedy12_99,
  author        = {{Kennedy}, G.~M. and {Wyatt}, M.~C. and {Sibthorpe}, B. and {Duch{\^e}ne}, G. and {Kalas}, P. and {Matthews}, B.~C. and {Greaves}, J.~S. and {Su}, K.~Y.~L. and {Fitzgerald}, M.~P.},
  title         = {{99 Herculis: host to a circumbinary polar-ring debris disc}},
  journal       = {\mnras},
  year          = 2012,
  month         = apr,
  volume        = {421},
  number        = {3},
  pages         = {2264-2276},
  doi           = {10.1111/j.1365-2966.2012.20448.x},
  archiveprefix = {arXiv},
  eprint        = {1201.1911},
  primaryclass  = {astro-ph.EP},
  adsurl        = {https://ui.adsabs.harvard.edu/abs/2012MNRAS.421.2264K}
}

@article{Kennedy19_circumbinary_protoplanetary,
  author  = {{Kennedy}, Grant M. and {Matr{\`a}}, Luca and {Facchini}, Stefano and {Milli}, Julien and {Pani{\'c}}, Olja and {Price}, Daniel and {Wilner}, David J. and {Wyatt}, Mark C. and {Yelverton}, Ben M.},
  title   = {{A circumbinary protoplanetary disk in a polar configuration}},
  journal = {Nature Astronomy},
  year    = 2019,
  month   = jan,
  volume  = {3},
  pages   = {230-235},
  doi     = {10.1038/s41550-018-0667-x},
  adsurl  = {https://ui.adsabs.harvard.edu/abs/2019NatAs...3..230K}
}

@article{Brinch16_misaligned_disks,
  author        = {{Brinch}, Christian and {J{\o}rgensen}, Jes K. and {Hogerheijde}, Michiel R. and {Nelson}, Richard P. and {Gressel}, Oliver},
  title         = {{Misaligned Disks in the Binary Protostar IRS 43}},
  journal       = {\apjl},
  year          = 2016,
  month         = oct,
  volume        = {830},
  number        = {1},
  eid           = {L16},
  pages         = {L16},
  doi           = {10.3847/2041-8205/830/1/L16},
  archiveprefix = {arXiv},
  eprint        = {1610.03626},
  primaryclass  = {astro-ph.SR},
  adsurl        = {https://ui.adsabs.harvard.edu/abs/2016ApJ...830L..16B}
}

@article{Lacour16_m-dwarf_star,
  author        = {{Lacour}, S. and {Biller}, B. and {Cheetham}, A. and {Greenbaum}, A. and {Pearce}, T. and {Marino}, S. and {Tuthill}, P. and {Pueyo}, L. and {Mamajek}, E.~E. and {Girard}, J.~H. and {Sivaramakrishnan}, A. and {Bonnefoy}, M. and {Baraffe}, I. and {Chauvin}, G. and {Olofsson}, J. and {Juhasz}, A. and {Benisty}, M. and {Pott}, J.-U. and {Sicilia-Aguilar}, A. and {Henning}, T. and {Cardwell}, A. and {Goodsell}, S. and {Graham}, J.~R. and {Hibon}, P. and {Ingraham}, P. and {Konopacky}, Q. and {Macintosh}, B. and {Oppenheimer}, R. and {Perrin}, M. and {Rantakyr{\"o}}, F. and {Sadakuni}, N. and {Thomas}, S.},
  title         = {{An M-dwarf star in the transition disk of Herbig HD 142527. Physical parameters and orbital elements}},
  journal       = {\aap},
  year          = 2016,
  month         = may,
  volume        = {590},
  eid           = {A90},
  pages         = {A90},
  doi           = {10.1051/0004-6361/201527863},
  archiveprefix = {arXiv},
  eprint        = {1511.09390},
  primaryclass  = {astro-ph.SR},
  adsurl        = {https://ui.adsabs.harvard.edu/abs/2016A&A...590A..90L}
}

@article{FernandezLopez17_strongly_misaligned,
  author        = {{Fern{\'a}ndez-L{\'o}pez}, M. and {Zapata}, L.~A. and {Gabbasov}, R.},
  title         = {{Strongly Misaligned Triple System in SR 24 Revealed by ALMA}},
  journal       = {\apj},
  year          = 2017,
  month         = aug,
  volume        = {845},
  number        = {1},
  eid           = {10},
  pages         = {10},
  doi           = {10.3847/1538-4357/aa7d51},
  archiveprefix = {arXiv},
  eprint        = {1707.01128},
  primaryclass  = {astro-ph.SR},
  adsurl        = {https://ui.adsabs.harvard.edu/abs/2017ApJ...845...10F}
}

@article{Kenworthy22_eclipse_v773,
  author        = {{Kenworthy}, M.~A. and {Gonz{\'a}lez Picos}, D. and {Elizondo}, E. and {Martin}, R.~G. and {van Dam}, D.~M. and {Rodriguez}, J.~E. and {Kennedy}, G.~M. and {Ginski}, C. and {Mugrauer}, M. and {Vogt}, N. and {Adam}, C. and {Oelkers}, R.~J.},
  title         = {{Eclipse of the V773 Tau B circumbinary disc}},
  journal       = {\aap},
  year          = 2022,
  month         = oct,
  volume        = {666},
  eid           = {A61},
  pages         = {A61},
  doi           = {10.1051/0004-6361/202243441},
  archiveprefix = {arXiv},
  eprint        = {2207.05575},
  primaryclass  = {astro-ph.SR},
  adsurl        = {https://ui.adsabs.harvard.edu/abs/2022A&A...666A..61K}
}

@article{Czekala19_degree_alignment,
  author        = {{Czekala}, Ian and {Chiang}, Eugene and {Andrews}, Sean M. and {Jensen}, Eric L.~N. and {Torres}, Guillermo and {Wilner}, David J. and {Stassun}, Keivan G. and {Macintosh}, Bruce},
  title         = {{The Degree of Alignment between Circumbinary Disks and Their Binary Hosts}},
  journal       = {\apj},
  year          = 2019,
  month         = sep,
  volume        = {883},
  number        = {1},
  eid           = {22},
  pages         = {22},
  doi           = {10.3847/1538-4357/ab287b},
  archiveprefix = {arXiv},
  eprint        = {1906.03269},
  primaryclass  = {astro-ph.EP},
  adsurl        = {https://ui.adsabs.harvard.edu/abs/2019ApJ...883...22C}
}

@article{Armstrong14_abundance_circumbinary,
  author        = {{Armstrong}, D.~J. and {Osborn}, H.~P. and {Brown}, D.~J.~A. and {Faedi}, F. and {G{\'o}mez Maqueo Chew}, Y. and {Martin}, D.~V. and {Pollacco}, D. and {Udry}, S.},
  title         = {{On the abundance of circumbinary planets}},
  journal       = {\mnras},
  year          = 2014,
  month         = oct,
  volume        = {444},
  number        = {2},
  pages         = {1873-1883},
  doi           = {10.1093/mnras/stu1570},
  archiveprefix = {arXiv},
  eprint        = {1404.5617},
  primaryclass  = {astro-ph.EP},
  adsurl        = {https://ui.adsabs.harvard.edu/abs/2014MNRAS.444.1873A}
}

@article{Nelder1965,
    author  = {Nelder, John A. and Mead, Roger},
    title   = {A Simplex Method for Function Minimization},
    journal = {The Computer Journal},
    volume  = {7},
    number  = {4},
    pages   = {308--313},
    year    = {1965},
    doi     = {10.1093/comjnl/7.4.308}
}

@ARTICLE{Virtanen2020_SciPy,
  author  = {Virtanen, Pauli and Gommers, Ralf and Oliphant, Travis E. and
             Haberland, Matt and Reddy, Tyler and Cournapeau, David and
             Burovski, Evgeni and Peterson, Pearu and Weckesser, Warren and
             Bright, Jonathan and {van der Walt}, St{\'e}fan J. and
             Brett, Matthew and Wilson, Joshua and Millman, K. Jarrod and
             Mayorov, Nikolay and Nelson, Andrew R. J. and Jones, Eric and
             Kern, Robert and Larson, Eric and Carey, C J and
             Polat, {\.I}lhan and Feng, Yu and Moore, Eric W. and
             {VanderPlas}, Jake and Laxalde, Denis and Perktold, Josef and
             Cimrman, Robert and Henriksen, Ian and Quintero, E. A. and
             Harris, Charles R. and Archibald, Anne M. and
             Ribeiro, Ant{\^o}nio H. and Pedregosa, Fabian and
             {van Mulbregt}, Paul and {SciPy 1.0 Contributors}},
  title   = {{{SciPy} 1.0: Fundamental Algorithms for Scientific
             Computing in Python}},
  journal = {Nature Methods},
  year    = {2020},
  volume  = {17},
  number  = {3},
  pages   = {261--272},
  doi     = {10.1038/s41592-019-0686-2}
}

@ARTICLE{Johnson2004_KH15D_binary,
       author = {{Johnson}, John Asher and {Marcy}, Geoffrey W. and {Hamilton}, Catrina M. and {Herbst}, William and {Johns-Krull}, Christopher M.},
        title = "{KH 15D: A Spectroscopic Binary}",
      journal = {\aj},
         year = 2004,
        month = sep,
       volume = {128},
       number = {3},
        pages = {1265-1272},
          doi = {10.1086/422735},
archivePrefix = {arXiv},
       eprint = {astro-ph/0403099},
 primaryClass = {astro-ph},
       adsurl = {https://ui.adsabs.harvard.edu/abs/2004AJ....128.1265J}
}

@article{Shappee2014_MAN,
  title      = {{{THE MAN BEHIND THE CURTAIN}}: {{X-RAYS DRIVE THE UV THROUGH NIR VARIABILITY IN THE}} 2013 {{ACTIVE GALACTIC NUCLEUS OUTBURST IN NGC}} 2617},
  shorttitle = {{{THE MAN BEHIND THE CURTAIN}}},
  author     = {Shappee, B. J. and Prieto, J. L. and Grupe, D. and Kochanek, C. S. and Stanek, K. Z. and De Rosa, G. and Mathur, S. and Zu, Y. and Peterson, B. M. and Pogge, R. W. and Komossa, S. and Im, M. and Jencson, J. and Holoien, T.W-S. and Basu, U. and Beacom, J. F. and Szczygie{\l}, D. M. and Brimacombe, J. and Adams, S. and Campillay, A. and Choi, C. and Contreras, C. and Dietrich, M. and Dubberley, M. and Elphick, M. and Foale, S. and Giustini, M. and Gonzalez, C. and Hawkins, E. and Howell, D. A. and Hsiao, E. Y. and Koss, M. and Leighly, K. M. and Morrell, N. and Mudd, D. and Mullins, D. and Nugent, J. M. and Parrent, J. and Phillips, M. M. and Pojmanski, G. and Rosing, W. and Ross, R. and Sand, D. and Terndrup, D. M. and Valenti, S. and Walker, Z. and Yoon, Y.},
  year       = 2014,
  month      = may,
  journal    = {The Astrophysical Journal},
  volume     = {788},
  number     = {1},
  pages      = {48},
  publisher  = {The American Astronomical Society},
  issn       = {0004-637X},
  doi        = {10.1088/0004-637X/788/1/48},
  urldate    = {2025-12-21},
  langid     = {english}
}

@article{Kochanek2017_All-Sky,
  title     = {The {{All-Sky Automated Survey}} for {{Supernovae}} ({{ASAS-SN}}) {{Light Curve Server}} v1.0},
  author    = {Kochanek, C. S. and Shappee, B. J. and Stanek, K. Z. and Holoien, T. W.-S. and Thompson, Todd A. and Prieto, J. L. and Dong, Subo and Shields, J. V. and Will, D. and Britt, C. and Perzanowski, D. and Pojma{\'n}ski, G.},
  year      = 2017,
  month     = aug,
  journal   = {Publications of the Astronomical Society of the Pacific},
  volume    = {129},
  number    = {980},
  pages     = {104502},
  publisher = {The Astronomical Society of the Pacific},
  issn      = {1538-3873},
  doi       = {10.1088/1538-3873/aa80d9},
  urldate   = {2025-12-21},
  langid    = {english}
}

@article{Fores-Toribio2025_ASASSN-24fw,
  author        = {{For{\'e}s-Toribio}, Raquel and {JoHantgen}, B. and {Kochanek}, C.~S. and {Jorstad}, S.~G. and {Hermes}, J.~J. and {Armstrong}, J.~D. and {Ashall}, C. and {Burns}, C.~R. and {Gaidos}, E. and {Hoogendam}, W.~B. and {Hsiao}, E.~Y. and {Medler}, K. and {Morrell}, N. and {Pfeffer}, C. and {Shappee}, B.~J. and {Stanek}, K. and {Tucker}, M.~A. and {Xiao}, H. and {Auchettl}, K. and {Lu}, L. and {Rowan}, D.~M. and {Vaccaro}, T. and {Williams}, J.~P.},
  title         = {{ASASSN-24fw: An 8-month long, 4.1 mag, optically achromatic and polarized dimming event}},
  journal       = {The Open Journal of Astrophysics},
  year          = 2025,
  month         = aug,
  volume        = {8},
  eid           = {114},
  pages         = {114},
  doi           = {10.33232/001c.143105},
  archiveprefix = {arXiv},
  eprint        = {2507.03080},
  primaryclass  = {astro-ph.SR},
  adsurl        = {https://ui.adsabs.harvard.edu/abs/2025OJAp....8E.114F}
}

@article{Zakamska2025_ASASSN-24fw,
  author        = {{Zakamska}, Nadia L. and {Adamane Pallathadka}, Gautham and {Bizyaev}, Dmitry and {Merc}, Jaroslav and {Owen}, James E. and {Reggiani}, Henrique and {Schlaufman}, Kevin C. and {B{\k{a}}kowska}, Karolina and {Bednarz}, S{\l}awomir and {Bernacki}, Krzysztof and {Gurgul}, Agnieszka and {Hall}, Kirsten R. and {Hambsch}, Franz-Josef and {Joachimczyk}, Barbara and {Kotysz}, Krzysztof and {Kurowski}, Sebastian and {Liakos}, Alexios and {Miko{\l}ajczyk}, Przemys{\l}aw J. and {Pak{\v{s}}tien{\.{e}}}, Erika and {Pojma{\'n}ski}, Grzegorz and {Popowicz}, Adam and {Reichart}, Daniel E. and {Wyrzykowski}, {\L}ukasz and {Zdanavi{\v{c}}ius}, Justas and {{\.Z}ejmo}, Micha{\l} and {Zieli{\'n}ski}, Pawe{\l} and {Zola}, Staszek},
  title         = {{ASASSN-24fw: Candidate circumplanetary disk occultation of a main-sequence star}},
  journal       = {arXiv e-prints},
  year          = 2025,
  month         = jul,
  eid           = {arXiv:2507.05367},
  pages         = {arXiv:2507.05367},
  doi           = {10.48550/arXiv.2507.05367},
  archiveprefix = {arXiv},
  eprint        = {2507.05367},
  primaryclass  = {astro-ph.EP},
  adsurl        = {https://ui.adsabs.harvard.edu/abs/2025arXiv250705367Z}
}

@article{astropy:2013,
  adsurl        = {http://adsabs.harvard.edu/abs/2013A%26A...558A..33A},
  archiveprefix = {arXiv},
  author        = {{Astropy Collaboration} and {Robitaille}, T.~P. and {Tollerud}, E.~J. and {Greenfield}, P. and {Droettboom}, M. and {Bray}, E. and {Aldcroft}, T. and {Davis}, M. and {Ginsburg}, A. and {Price-Whelan}, A.~M. and {Kerzendorf}, W.~E. and {Conley}, A. and {Crighton}, N. and {Barbary}, K. and {Muna}, D. and {Ferguson}, H. and {Grollier}, F. and {Parikh}, M.~M. and {Nair}, P.~H. and {Unther}, H.~M. and {Deil}, C. and {Woillez}, J. and {Conseil}, S. and {Kramer}, R. and {Turner}, J.~E.~H. and {Singer}, L. and {Fox}, R. and {Weaver}, B.~A. and {Zabalza}, V. and {Edwards}, Z.~I. and {Azalee Bostroem}, K. and {Burke}, D.~J. and {Casey}, A.~R. and {Crawford}, S.~M. and {Dencheva}, N. and {Ely}, J. and {Jenness}, T. and {Labrie}, K. and {Lim}, P.~L. and {Pierfederici}, F. and {Pontzen}, A. and {Ptak}, A. and {Refsdal}, B. and {Servillat}, M. and {Streicher}, O.},
  doi           = {10.1051/0004-6361/201322068},
  eid           = {A33},
  eprint        = {1307.6212},
  journal       = {\aap},
  month         = oct,
  pages         = {A33},
  primaryclass  = {astro-ph.IM},
  title         = {{Astropy: A community Python package for astronomy}},
  volume        = 558,
  year          = 2013
}

@ARTICLE{Schechter1993_DoPHOT,
       author = {{Schechter}, Paul L. and {Mateo}, Mario and {Saha}, Abhijit},
        title = "{DoPHOT, A CCD Photometry Program: Description and Tests}",
      journal = {\pasp},
         year = 1993,
        month = nov,
       volume = {105},
        pages = {1342},
          doi = {10.1086/133316},
       adsurl = {https://ui.adsabs.harvard.edu/abs/1993PASP..105.1342S}
}

@ARTICLE{Bouchy2001,
       author = {{Bouchy}, F. and {Pepe}, F. and {Queloz}, D.},
        title = "{Fundamental photon noise limit to radial velocity measurements}",
      journal = {\aap},
         year = 2001,
        month = aug,
       volume = {374},
        pages = {733-739},
          doi = {10.1051/0004-6361:20010730},
       adsurl = {https://ui.adsabs.harvard.edu/abs/2001A&A...374..733B}
}
\bibliographystyle{aasjournalv7}

\end{document}